\documentclass[]{aastex631}
\usepackage{amsmath}
\usepackage{amssymb}
\usepackage{graphicx}
\usepackage{hyperref}
\submitjournal{ApJ}

\newcommand*\mean[1]{\overline{#1}}

\defcitealias{icru1980}{ICRU 33 (1980)}

\shorttitle{Impact of CRs on the Atmosphere of TRAPPIST-1e}
\shortauthors{Herbst et al.}

\begin{document}
\title{Impact of Cosmic Rays on Atmospheric Ion Chemistry and Spectral Transmission Features of TRAPPIST-1e}

\author[0000-0001-5622-4829]{Konstantin Herbst}\affiliation{Institut f\"ur Experimentelle und Angewandte Physik, Christian-Albrechts-Universit\"at zu Kiel, 24118 Kiel, Germany}

\author[0000-0002-8631-7572]{Andreas Bartenschlager}
\affiliation{Institut f\"ur Meteorologie und Klimaforschung, Karlsruher Institut für Technologie, 76344 Eggenstein-Leopoldshafen, Germany}

\author[0000-0003-3646-5339]{John Lee Grenfell}
\affiliation{Institut f\"ur Planetenforschung, Deutsches Zentrum f\"ur Luft- und Raumfahrt  (DLR), 12489 Berlin, Germany}

\author[0000-0003-2329-418X]{Nicolas Iro}
\affiliation{Institut f\"ur Planetenforschung, Deutsches Zentrum f\"ur Luft- und Raumfahrt  (DLR), 12489 Berlin, Germany}

\author[0000-0002-3527-9051]{Miriam Sinnhuber}
\affiliation{Institut f\"ur Meteorologie und Klimaforschung, Karlsruher Institut für Technologie, 76344 Eggenstein-Leopoldshafen, Germany}

\author[0000-0002-0856-4340]{Benjamin Taysum}
\affiliation{Institut f\"ur Planetenforschung, Deutsches Zentrum f\"ur Luft- und Raumfahrt  (DLR), 12489 Berlin, Germany}

\author[0000-0002-2238-5269]{Fabian Wunderlich}
\affiliation{Institut f\"ur Planetenforschung, Deutsches Zentrum f\"ur Luft- und Raumfahrt  (DLR), 12489 Berlin, Germany}

\author[0000-0003-3659-7956]{N. Eugene Engelbrecht}
\affiliation{Centre for Space Research, North-West University, 2520, Potchefstroom, South Africa}

\author[0000-0002-9788-5540]{Juandre Light}
\affiliation{Centre for Space Research, North-West University, 2520, Potchefstroom, South Africa}
\affiliation{South African National Space Agency, Hermanus, South Africa}

\author[0000-0002-4840-6355]{Katlego D. Moloto}
\affiliation{Centre for Space Research, North-West University, 2520, Potchefstroom, South Africa}

\author[0000-0001-8935-2472]{Jan-Vincent Harre}
\affiliation{Institut f\"ur Planetenforschung, Deutsches Zentrum f\"ur Luft- und Raumfahrt (DLR), 12489 Berlin, Germany}

\author[0000-0002-6510-1828]{Heike Rauer}
\affiliation{Institut f\"ur Planetenforschung, Deutsches Zentrum f\"ur Luft- und Raumfahrt (DLR), 12489 Berlin, Germany}
\affiliation{Institut f\"ur Geologische Wissenschaften, Freie Universit\"at Berlin, 12249 Berlin, Germany}

\author[0000-0001-7196-6599]{Franz Schreier}
\affiliation{Institut f\"ur Methodik der Fernerkundung, Deutsches Zentrum f\"ur Luft- und Raumfahrt (DLR), 82234 Oberpfaffenhofen, Germany}

\correspondingauthor{Konstantin Herbst}
\email{herbst@physik.uni-kiel.de}

\begin{abstract}
\hspace{-0.3cm}\small{Ongoing observing projects like the James Webb Space Telescope (JWST) and future missions offer the chance to characterize Earth-like exoplanetary atmospheres. Thereby, M-dwarfs are preferred targets for transit observations, for example, due to their favorable planet-star contrast ratio. However, the radiation and particle environment of these cool stars could be far more extreme than what we know from the Sun. Thus, knowing the stellar radiation and particle environment and its possible influence on detectable biosignatures - particularly signs of life like ozone and methane - is crucial to understanding upcoming transit spectra. In this study, with the help of our unique model suite INCREASE, we investigate the impact of a strong stellar energetic particle event on the atmospheric ionization, neutral and ion chemistry, and atmospheric biosignatures of TRAPPIST-1e. Therefore, transit spectra for six scenarios are simulated. We find that a Carrington-like event drastically increases atmospheric ionization and induces substantial changes in ion chemistry and spectral transmission features: all scenarios show high event-induced amounts of nitrogen dioxide (i.e., at 6.2 $\mu$m), a reduction of the atmospheric transit depth in all water bands (i.e., at 5.5 -- 7.0 $\mu$m), a decrease of the methane bands (i.e., at 3.0 -- 3.5 $\mu$m), and depletion of ozone (i.e., at $\sim$ 9.6 $\mu$m). Therefore, it is essential to include high-energy particle effects to correctly assign biosignature signals from, e.g., ozone and methane. We further show that the nitric acid feature at 11.0 - 12.0 $\mu$m, discussed as a proxy for stellar particle contamination, is absent in wet-dead atmospheres.
} 
\end{abstract}


\section{Introduction}\label{sec:intro}

More than 5500 exoplanets in over 4100 exoplanetary systems\footnote{see, e.g., \url{https://exoplanets.nasa.gov}, last visited on October 13, 2023} have been detected since the discovery of the first Jupiter-mass exoplanet around a Sun-like star in 1995 \citep{Mayor-Queloz-1995}; ranging from hot gas giants to rocky Earth-like exoplanets around much smaller and cooler stars (i.e., M-dwarfs).  According to \citet{Hill-etal-2023}, as of today, 17 of the rocky exoplanets with radii up to 1.6 R$_{\earth}$, or masses up to 3 M$_\earth$, are believed to lie in the conservative habitable zone (HZ), a region around the star determined by the runaway greenhouse in which the stellar radiation from one or multiple host stars allows for liquid water to exist for geological periods on the surface of the orbiting rocky exoplanets \citep[e.g.,][]{Kasting-etal-1993}.

Within the next decade, next-generation telescopes will provide new opportunities to study the atmospheres of Earth-like exoplanets. Currently, particularly M-dwarfs (making up to 75\% of all stars in the solar neighborhood) are favored targets for transit observations due to their favorable planet-star contrast ratio and shorter orbital periods (more transit events over a given time). 

Currently, one of the most exciting systems is that of TRAPPIST-1. The nearby ultracool M-dwarf has several Earth-sized exoplanets, three of which, i.e., planets e (at 0.028 AU), f (at 0.038 AU), and g (at 0.047 AU), are assumed to lie within the conservative stellar HZ \citep{Hill-etal-2023}, and thus may have equilibrium temperatures that support liquid surface water, one of the key ingredients for life as we know it from Earth. The first JWST observations of TRAPPIST-1b showed no atmospheric absorption of any species \citep{GreeneEA23, IhEA23}, while, with the help of JWST/NIRISS/SOSS transmission spectra, \citet{LimEA23} found strong stellar contamination in the order of hundreds of ppm and ruled out H$_2$-rich atmospheres. Thus, distinguishing between TRAPPIST-1b being a bare rock and/or having a thin atmosphere is currently impossible. 
As discussed by \citet{ZiebaEA23}, a thin, O$_2$-dominated, low CO$_2$ atmosphere or a bare rock surface are possible explanations for the secondary eclipse depth of 421$\pm$94 ppm at 15 $\mu$m that was revealed by the first JWST observations of TRAPPIST-1 c.
Most recently, however, \citet{LincowskiEA23} showed that steam atmospheres of $\ge$ 0.1 bar and - although less likely - thick O$_2$-dominated atmospheres are also consistent with the observations. Nevertheless, our current knowledge of the TRAPPIST-1 system and its evolution does not rule out the possibility of Earth-like atmospheres on the planets further out in the system, particularly those within the HZ \citep[see, e.g., discussion in][]{Krissansen-Totton2023}.

 Nevertheless, TRAPPIST-1 is known to be an active star with frequent (high-energy) stellar flaring \citep[e.g.,][]{Vida-etal-2017} and EUV/XUV activity \citep[][]{Wheatley-etal-2017}, which could strip away the planetary atmosphere and impact its habitability \citep[e.g.,][]{Bourrier-etal-2017} by, e.g., reducing the chances for life to develop and persist on the planets within the HZ. As shown by, e.g., \citet{Herbst-etal-2019c}, \citet{Scheucher-etal-2020a}, and \citet{Barth-etal-2021} in particular, cosmic rays of galactic and stellar origin play a crucial role in determining the atmospheric climate and chemistry and habitability on (Earth-like) exoplanets. Consistent studies on the impact of cosmic rays, however, are missing.

Thus, at the dawn of the age of atmospheric characterization of such Earth-like exoplanets with JWST, it is of utmost importance to study the atmospheric response to the stellar particle and radiation environment as preparation for interpreting the observations soon to come. By utilizing the INCREASE model suite, we - for the first time - will shed light on the impact of cosmic ray effects on the ion-chemistry of a potentially Earth-like TRAPPIST-1e atmosphere.

\section{Scientific Background}\label{sec:SciBgr}

\subsection{The Galactic Cosmic Ray Background}\label{sec:GCRs}
The transport of GCRs within stellar astrospheres depends on different factors, such as the stellar type of the host star, its rotation rate, the stellar wind dynamics, and the stellar activity. Thereby, the stellar magnetic field defines the inner boundary conditions, while the local interstellar spectrum (LIS) forms the outer boundary condition. First model studies by \citet{Herbst-etal-2020a} have shown that cool star astrospheres may be rather diverse.

Based on our heliospheric knowledge, the modulation of GCRs within cool star astrospheres is usually based on solving the transport equation by \citet{Parker-1965} either analytically with the help of the Force Field solution \citep[FFs][]{Gleeson-Axford-1968} or numerically by, e.g., employing stochastic differential equations \citep[SDEs, see][and references therein]{Strauss-Effenberger-2017, MolotoEA19}. 

However, little is known about the astrospheric environments of cool stars. Thus, either a modified FFs \citep[e.g.,][]{Mesquita-etal-2021, Rodgers-Lee-etal-2021} or a 1D version of the transport equation are used to solve the transport of GCRs within astrospheres \citep[e.g., ][]{Herbst-etal-2020b, Mesquita-etal-2022}. By employing results from astrospheric 3D MHD modeling, \citet{Herbst-etal-2020b} suggested for the first time significant differences between the analytic and numerical solution, emphasizing that the GCR flux, particularly for cool star astrospheres, might have a much more significant impact on exoplanetary atmospheres, and thus their habitability, than previously thought. This is not unexpected, given the limitations implicit to the analytical Force Field and Convection-Diffusion approaches to solving Parker's transport equation \cite[TPE, see, e.g.,][]{CM04, CaballeroEA19, EngelbrechtdiFelice20}. \citet{Herbst-etal-2020b} further concluded that 3D transport modeling is mandatory to properly describe the GCR transport, which, however, is difficult given the relatively unknown behavior of the astrospheric plasma environments \cite[see also][]{LightEA22}. More information on the importance of turbulence in astrospheres, known to be essential in modeling GCR transport from first principles \citep[see also][]{EngelbrechtEA22} are discussed in \citet{Herbst-etal-2022}. Here, we solve the 1D Parker transport equation following the SDE approach outlined by \citet{EngelbrechtdiFelice20} and \citet{LightEA22}, where
\begin{equation}\begin{split}
    \frac{\partial{f}}{\partial{t}} &= \frac{1}{r^2}\frac{\partial}{\partial{r}}(r^2\kappa_{rr}\frac{\partial{f}}{\partial{r}}) + \frac{1}{3r^2}\frac{\partial}{\partial{r}}(r^2V_s)\frac{1}{p^2}\frac{\partial}{\partial{p}}(p^3f)\\
    &- \frac{1}{r^2}\frac{\partial}{\partial{r}}(r^2V_sf),
\end{split}\end{equation}\label{TPE}
expressed in terms of the omnidirectional GCR distribution function $f$, which is related via the momentum $p$ to differential intensity by $f(r,p,t)=p^{-2}j_{T}$ \cite[see, e.g.,][]{Moraal13}. The above equation models GCR adiabatic energy changes, convection via the stellar wind speed $V_s$, and radial diffusion, controlled by the diffusion coefficient $\kappa_{rr}$. This quantity is modeled using the approach of \citet{LightEA22}, which employs an essentially Bohm diffusion coefficient in terms of its magnetic field dependence \cite[see, e.g.,][]{Shalchi09}, but modified with the rigidity dependence expected of a perpendicular diffusion coefficient derived using the nonlinear guiding center theory of \citet{MattEA03-diff}, given by
\begin{equation}
    \kappa_{rr} = \frac{v}{3}\lambda_{0}\left(\frac{B_0}{B}\right)\left(\frac{P}{P_0}\right)^{2/3}\label{eq:lambdaperp}
\end{equation}
with $P$ the particle rigidity, and $P_0=1$~GV. The astrospheric magnetic field (AMF) is denoted by $B$ and normalized to a value of $B_0$ at a distance $r_0 =1$~AU. The mean free path is scaled with the parameter $\lambda_0=0.01$~AU to a value at $r_0$ within the range of what is expected from heliospheric modulation studies \cite[see, e.g.,][and references therein]{EngelbrechtEA22}.

Several assumptions must be made to model the large-scale plasma quantities required to solve the Parker TPE. Firstly, it is assumed that the AMF of TRAPPIST-1 can be described using a \citet{parker1958} magnetic field model, at least within its termination shock. This assumption is based on the typical behavior of MHD-simulated AMFs for this type of star in this region \cite[see][]{Herbst-etal-2022}. To calculate such a field magnitude, we assume a stellar rotation period of $3.3$~days \citep{LugerEA17}, a stellar wind velocity of $1400$~ km/s \citep{GarraffoEA17}, and a stellar surface field of $600$~G \cite[][and references therein]{HarbachEA21}. Accordingly, we assume a value of $1.8$~nT for the AMF magnitude at $1$~AU. Given that, to the best of our knowledge, no large-scale MHD simulations for the TRAPPIST-1 astrosphere have been published, it is not possible to employ an estimate for its termination shock location based on the limited studies available. Therefore, it is assumed here as a first approach that its termination shock is located at $76$~AU, following the results of \citet{Herbst-etal-2020b} for Proxima Centauri. Given that the mass-loss rate for TRAPPIST-1 is expected to be larger than that for Proxima Centauri \citep{GarraffoEA17}, this estimate is probably too low and may need to be revisited when full large-scale MHD simulations for the astrosphere become available in the literature. Since such an analytical approach is impossible for the astrosheath, we employ GCR proton and helium boundary spectra at this termination shock expression when we solve Eq.~(\ref{TPE}). These are modeled following the approach of \citet{EngelbrechtMoloto21}, who fit spectral forms to Voyager observations for galactic cosmic ray intensities reported by \citet{WebberEA08}. For GCR protons, the boundary spectrum is given by 
\begin{equation}
j_{B}(76~\mathrm{AU, H})=\frac{17.0 (P/P_{o})^{-2.4}}{2.2+2.1(P/P_{o})^{-3}},
\label{eq:lisT}
\end{equation}
with units of particles m$^{-2}$ s$^{-1}$ sr$^{-1}$ MeV$^{-1}$, where $P$ is in units of GV, and $P_o$ is chosen to be equal to $1$~GV. Note that here no time dependence is assumed for Eq.~(\ref{eq:lisT}), as was done by \citet{EngelbrechtMoloto21}. 

The transport of helium within astrospheres has not previously been considered, even though it has been relatively well-studied in the heliospheric context \cite[see, e.g.][]{ShenEA19}. In this study, we assume the helium spectrum at 76 AU to be given by 
\begin{equation}
j_{B}(76~\mathrm{AU, He})=\frac{3.3 (P/P_{o})^{-2.5}}{(1.2+5.0(P/P_{o})^{-4.4})^{0.92}},
\label{eq:lisT2}
\end{equation}
constructed in the same manner as was Eq.~\ref{eq:lisT}, and will use the 1D SDE code at hand to study the modulation of these particles for the first time in an astrospheric context.

\begin{figure}[!t]
\vspace{0.3cm}
    \centering
    \includegraphics[width=0.7\columnwidth]{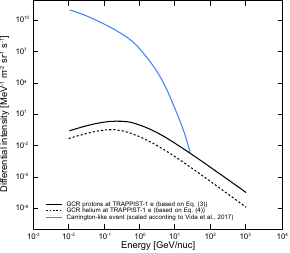}
    \caption{Differential primary GCR spectra for a quiescent TRAPPIST-1 (black lines) in comparison to a Carrington-like event scaled to the location of TRAPPIST-1e \citep[see also][]{Vida-etal-2017}.}
    \label{fig:2}
\end{figure}

Although Eqs.~(\ref{eq:lisT}) and~(\ref{eq:lisT2}) are strictly only applicable to heliospheric conditions, they nevertheless give an estimate of the modulation effects due to the astrosheath in that they represent a considerable reduction of GCR intensities as expected in the local interstellar medium \cite[see, e.g.,][]{Herbst-etal-2017}. This is because considerable GCR modulation has been reported in the heliosheath \cite[e.g.][]{StoneEA13}, and the same could be reasonably expected for astrospheres. The above expressions are shown, along with the calculated differential intensity spectra at the position of TRAPPIST-1e in Fig.~\ref{fig:2}. Both spectra show a relatively low degree of modulation compared to Earth, with most modulation occurring at energies below $\sim 1$~Gev/nucleon. This behavior is similar to that reported by \citet{Herbst-etal-2020b} for Proxima Centauri b, resulting from the AMF dependence of the radial diffusion coefficient. At higher energies, the spectra remain almost unmodulated, as would be expected from the rigidity dependence of Eq.~(\ref{eq:lambdaperp}). 

\subsection{The Stellar Particle Environment}\label{sec:SEPspectra} 
One of the strongest solar flares ever observed in the heliosphere is the Carrington event, which occurred in September 1859. The corresponding flare was estimated to have had an energy of 10$^{33}$ erg \citep{Cliver-etal-2022} and resulted in one of the strongest geomagnetic storms ever impacting the Earth's atmosphere. Cosmogenic radionuclide records, in particular $^{10}$Be, $^{14}$C, and $^{36}$Cl, however, suggest short-period increases in the production rates about 20 times larger than changes ascribed to solar modulation, indicating the existence of much more extreme events in the past \citep[e.g., AD774/775][]{Miyake-etal-2012, Mekhaldi-etal-2015} that lead to extreme particle events strongly impacting the Earth's atmosphere. Unfortunately, information on the corresponding event spectra of historical events {such as the} Carrington event is lacking. For example, to this date, it is scientifically debated whether the corresponding production rate increases around AD774/775 were caused by a single extreme event or by multiple strong events within a short amount of time \citep[e.g.,][]{Cliver-etal-2022, Papaioannou-etal-2023}.  Thus, historical events are usually scaled with the help of measured modern-era Ground Level Enhancement (GLE) events - strong solar energetic particle events that are detected by neutron monitors at the Earth's surface - on average occurring once every year.

\citet{Vida-etal-2017} analyzed the 80-day continuous short-cadence K2 observations of TRAPPIST-1 and revealed 42 flare events with integrated energies between 1.26$\times$10$^{30}$ and 1.24$\times 10^{33}$ erg. Thus, Carrington-like events on TRAPPIST-1 occur frequently. As a first step, in this study, we investigate the impact of a single strong event. As suggested by \citet{Vida-etal-2017}, therefore, the terrestrial Carrington event - based on the spectral shape of GLE44 (October 1989), with a rather flat spectrum but much higher intensities at lower energies - has been scaled to the location of TRAPPIST-1e (0.029 AU) via $1/r^2$ (see blue line in Fig.~\ref{fig:2}), which makes the Carrington-like event at TRAPPIST-1e more than three orders of magnitude stronger than the actual Carrington event that impacted the Earth. Further, we assume that particles - in principle - can be accelerated to higher energies during such strong flares and allowed particle energies up to 40 GeV.  Note that according to the flare frequency distribution \citep[e.g.,][]{Vida-etal-2017, Glazier-etal-2020} TRAPPIST-1 is a constantly flaring star, indicating that Carrington-like flares could occur once every three months and that much stronger events could occur on an annual and decennial basis. Thus, the stellar activity of TRAPPIST-1 could constantly impact the atmospheres of its planets. 
In this study, we focus on the impact of a single Carrington-like event, while the impact of constant flaring will be discussed in a follow-up paper.  Furthermore, note that even stronger events might occur at TRAPPIST-1e \citep[according to ][up to six orders of magnitude difference to current solar events]{Fraschetti-etal-2019} while it might also be possible that the stellar magnetic field of M-dwarfs may prevent energetic particles from escaping during stellar flares and that corresponding CMEs may be confined \citep[see][]{Alvarado-Gomez-2019, Fraschetti-etal-2019}.
\subsection{Cosmic Ray Induced Atmospheric Ion Pair Production}\label{sec:IPPR}
CRs entering a planetary atmosphere mainly lose their energy due to the ionization of ambient matter. Consequently, once the nucleonic-electromagnetic particle shower reaches the surface, the chances of it interacting with the surrounding gases (and the planetary surface) dramatically increase. Ionization of the lower and middle atmosphere occurs because of the production of charged secondary particles within the electromagnetic branch. The severity of this effect, however, varies considerably depending on several factors, including the energy of the primary particle, the type of the produced secondary particle, and the atmospheric depth \citep[e.g.,][]{Banjac-Herbst-Heber-2019}.

While the GCR-induced ionization is likely to be anti-correlated to the stellar activity (if present), a stellar energetic particle event can significantly contribute to atmospheric ionization only if very high-energetic particles are produced in stellar flares and coronal mass ejections (CMEs). 

The CR-induced ionization rate $Q$ as a function of the stellar activity (via the stellar modulation parameter $\phi$), cutoff energy ($E_C$), and the atmospheric altitude $x$ numerically is described by
\begin{equation}
    Q(\phi, E_C, x) = \sum_i \int_{E_{C, i}}^{\infty} J_i(\phi, E) \cdot Y_i(E,x)~dE,
\end{equation}
where $i$ represents the type of the primary cosmic ray particle (here protons and $\alpha$-particles), $J_i(\phi, E)$ the primary differential energy spectrum of either GCRs or stellar energetic particles and $Y_i(E,x)$ the atmospheric ionization yield. 

The latter can be computed as 
\begin{equation}
    Y_i(E,x) = \alpha \cdot \frac{1}{E_{ion}} \cdot \frac{\Delta E_i}{\Delta x},
\end{equation}
and thus depends on the geometrical normalization factor $\alpha$ = 2$\pi \int \cos(\theta) \sin{\theta}~d\theta$, the depth-dependent mean specific energy loss $\frac{\Delta E_i}{\Delta x}$, and the atmospheric ionization energy $E_{ion}$. Since in all the investigated scenarios, Earth-like atmospheres - defined here as N$_2$-O$_2$-CO$_2$ dominated with 1-2 bar surface pressure (see Sec.~\ref{sec:Scenarios}) - are being studied, an average ionization energy of 31.7 eV \citep[see, e.g.,][]{Wedlund-etal-2011} is used, and an Earth-like magnetic field is assumed.
\subsection{Cosmic Ray Induced Atmospheric Radiation Exposure}
Besides atmospheric ionization profiles, also the atmospheric radiation exposure can be modeled with codes such as the Atmospheric Radiation Interaction Simulator \citep[AtRIS,][see Sec.~\ref{sec:AtRIS}]{Banjac-Herbst-Heber-2019}. As discussed in \citet{Herbst-etal-2020a}, the pre-calculated relative ionization efficiency $\mathcal{I}_{R,j}$ given as $\mathcal{I}_{R,j} \left(E_i\right):= \frac{E_d}{E_i}$, with $E_{d}$ representing the average ionization energy a particle of type $j$ is causing in a well-defined phantom like the phantom of the International Commission on Radiation Units and Measurement (ICRU) \citep[e.g.,][mimicking the human body]{ICRU-1981}, is used. The average absorbed dose of the ICRU phantom is given as
\begin{equation}
        \mean{D_j}\left(E_i, r\right) = \mathcal{I}_{R,j}(E_i) \cdot \frac{E_i}{m_{\mathrm{ph}}},
    \end{equation}
where E$_i$ is the ionization energy and $m_{\mathrm{ph}}$ the mass of the phantom given by \mbox{$\rho \cdot \frac{4}{3}\pi \cdot r_{\mathrm{ph}}^3$}. Finally, the results are convoluted with the primary particle spectrum and summed up over all energy bins and particle types (i.e., protons and alphas).

\subsection{Cosmic Ray Induced Ion Chemistry}
Formation of NOx -- The dissociation of N$_2$ by charged particle impact and fast ion chemistry reactions can lead to the production of NOx species (N, NO, NO$_2$) \citep[see, e.g.,][and references therein for recent reviews of the terrestrial atmosphere]{Sinnhuber-etal-2012, Sinnhuber2019}. While positively charged ions dominate in the lower thermosphere, negatively charged ions become more common in the layers below, affecting the partitioning and lifetime of NOx species. Recombination of NO$^+$ and N$_2^+$ with electrons is an important source of NOx. In addition, charge transfer of N$^+$ \citep{Nicolet1975} and ion-neutral reactions also contribute to the formation of NOx \citep{Nicolet1965}. The nitrogen atoms formed by recombination can also occur in energetically different states. On the one hand, in the ground state N($^4S$), on the other hand, also in the excited states N($^2D$) and N($^2P$). They can react with O$_2$ and O$_3$ to form NO \citep{Sinnhuber-etal-2012}. The formation of nitrogen oxides NO and NO$_2$ as a consequence of particle precipitation is well known on Earth but has been found to depend on the large availability of N$_2$ as a nitrogen source. The resulting NOx formation caused by the Carrington-like event for the different scenarios is shown in the upper left panel of Fig.~\ref{fig:14}(b). As nitrogen is one of the most abundant species in the CO$_2$-poor scenarios, large amounts of NOx are formed with amounts comparable to large particle events in the Earth's atmosphere \citep[e.g.,][]{Jackman2000, Jackman2001, Funke2011}.

Production of HO$_x$ and loss of water vapor -- Water vapor is taken up into positive ions, forming large positive water cluster ions. In the process of water vapor uptake and in the recombination of the cluster ions with electrons or negative ions, H and OH are released into the gas phase. This process of HO$_x$ formation was postulated for the first time based on observations of ozone loss by \cite{SwiderKeneshea1973} and formulated concisely by \cite{Solomon1981}. H and OH are released in equal parts, and this HO$_x$ production is balanced by a loss of one molecule of water vapor per H+OH pair. A summary of the reaction chains is provided, e.g., in \cite{Sinnhuber-etal-2012}.

Changes in Ozone -- \cite{Bates1950}, for the first time, formulated that HOx catalytic cycles could constitute a key ozone loss mechanism. Similar mechanisms of catalytic ozone loss have been discussed for the terrestrial atmosphere and have been postulated for NOx \cite[e.g.,][]{Crutzen1970}.
\begin{figure*}[!t]
    \centering
    \includegraphics[width=0.9\textwidth]{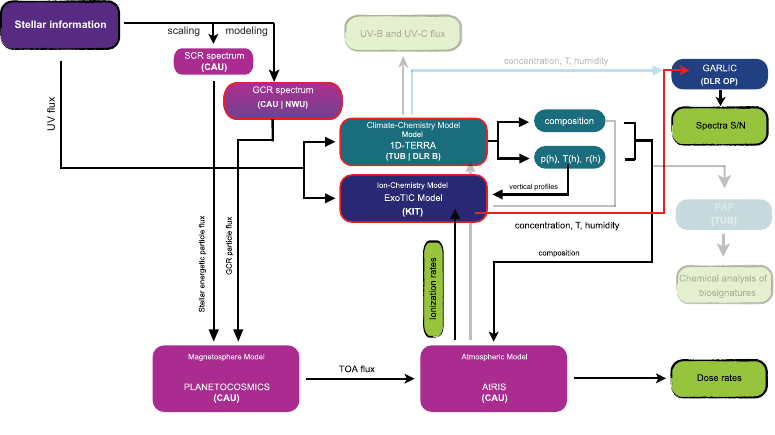}
    \caption{Interplay and output of the modified INCREASE model suite to study the impact of CRs on the atmospheric ion-chemistry and induced changes to the spectral transmission features. Purple boxes highlight components derived and utilized at the University of Kiel (CAU). In contrast, dark blue boxes represent the models utilized by DLR Berlin and the Technical University Berlin (DLR/TUB). The light blue box represents the \textit{ExoTIC} model maintained at the Karlsruher Institute of Technology (KIT), and the blue box highlights the \textit{GARLIC} model maintained at DLR Oberpfaffenhofen. Green boxes show the output provided by the modified model suite. Unused branches and models of the full model suite \citep[][]{Herbst-etal-2019c} are shown transparently; blocks with red outlines indicate updated models/codes; red arrows indicate newly added modeling steps.}
    \label{fig:1}
\end{figure*}
\section{The modified INCREASE Model Suite}\label{sec:INCREASE}
An interpretation of upcoming atmospheric exoplanetary observations will need dedicated model studies taking into account processes such as the transport of stellar energetic particles (StEPs) and galactic cosmic rays (GCRs) through stellar astrospheres, planetary magnetic fields, and exoplanetary atmospheres, atmospheric escape, outgassing, climate, and (photo/ion) chemistry.

As a first step, \citet{Herbst-etal-2019c, Herbst-etal-2021} built up the model suite INCREASE, a simulation chain that couples state-of-the-art magnetospheric and atmospheric propagation and interaction models \textit{PLANETOCOSMICS} \citep{desorgher2006planetocosmics} and \textit{AtRIS} \citep{Banjac-Herbst-Heber-2019} with the atmospheric chemistry and climate model \textit{1D-TERRA} \citep[e.g.,][]{Wunderlich-etal-2020, Scheucher-etal-2020b} and the ion chemistry model \textit{ExoTIC} \citep[based on UBIC, see, e.g., ][]{Sinnhuber-etal-2012}. 

This study focuses on the impact of cosmic rays (CRs) on atmospheric ionization, ionization-induced changes in ion chemistry, and spectral transmission features. As shown in  Fig.~\ref{fig:1}, we slightly modified our model suite. Note that the parts of the full model suite described in \citet{Herbst-etal-2019c} not used in this study are displayed transparently. Updated codes and models are highlighted by red outlining, while red arrows indicate newly added modeling steps. A brief description of the codes maintained at our institutes and the model coupling within the INCREASE model suite is given in what follows. 
\subsection{AtRIS}\label{sec:AtRIS}
The Atmospheric Radiation Interaction Simulator \citep[\textit{AtRIS}, ][]{Banjac-Herbst-Heber-2019} is a GEANT4-based \citep[][]{Agostinelli-etal-2003} code to model the particle transport and interaction within exoplanetary atmospheres. We use \textit{AtRIS} to model the cosmic ray-induced atmospheric ionization rates and radiation dose profiles of diverse exoplanetary atmospheres.
\subsection{PLANETOCOSMICS}
\textit{AtRIS} does not incorporate (exo)planetary magnetic fields. The location and altitude-dependent cutoff rigidity values — measures of the energy a particle must have to reach a specific location within the (exo)planetary atmosphere — are therefore simulated using PLANETOCOSMICS \citep{desorgher2006planetocosmics}. The particle fluxes atop the exoplanetary atmosphere are estimated depending on the computed values.  
\subsection{1D-TERRA}
1D-TUB model used in \citet{Herbst-etal-2019c} has been considerably modified to the 1D Terrestrial Climate-Chemistry (1D-TERRA) column model \citep{Wunderlich-etal-2020, Scheucher-etal-2020b}. The radiative scheme was extensively revised to be valid for a wide range of exoplanetary atmospheres up to 1000~K and 1000~bar \citep[e.g.,][]{Scheucher-etal-2020a}. In the visible and infrared, up to 20 absorbers can be chosen, and overall, up to 81 UV/visible cross-sections can be added. Rayleigh scattering and various continua can be added flexibly.
 
\citet{Wunderlich-etal-2020} further updated the flexible chemical network, which currently consists of 1127 reactions for 115 species. The scheme can consider wet and dry deposition and biomass, volcanic, and lightning emissions and features an adaptive eddy-diffusion coefficient profile based on atmospheric conditions.
\subsection{ExoTIC} 
Since an ion chemistry network is not explicitly included in \textit{1D-TERRA} to study the effects of cosmic rays on atmospheric ion chemistry, we further coupled our models to \textit{ExoTIC}, an adapted version of the University of Bremen Ion Chemistry column model \citep[UBIC, ][]{Winkler-etal-2009, Sinnhuber-etal-2012, Nieder-etal-2014}. \textit{ExoTIC} is a 1D stacked box model of the neutral and ion atmospheric composition. As of today, \textit{ExoTIC} considers 60 neutral and 103 charged species (54 positive and 49 negative - including electrons), which interact due to neutral, neutral-ion, ion-ion gas-phase reactions, photolysis, and photo-electron attachment and detachment reactions \citep[see, e.g.,][]{Sinnhuber-etal-2012}. So far, \textit{ExoTIC} has been extended to model (rocky) exoplanets with N$_2$-O$_2$ and CO$_2$-dominated atmospheres. Time-dependent photochemistry is driven by time-variable stellar flux based on the stellar spectrum and the planet's orbital motion and rotation. Particle impact ionization is considered a starting point for ion chemistry by dissociation, dissociative ionization, and ionization of $N_2$, $O_2$, and $CO_2$ as a function of time-dependent atmospheric ionization. Since the ions are much shorter-lived than the neutral species, the ion composition is calculated every six minutes as photochemical equilibrium, and formation/loss rates of the neutral species due to the ion composition are calculated and transferred to the neutral chemistry scheme.
Since \citet{Herbst-etal-2019c}, the particle impact ionization, dissociative ionization and ionization of $CO_2$ has been added 
\begin{eqnarray}
CO_2 + e* \longrightarrow CO_2^+ + e^-\\
CO_2 + e* \longrightarrow CO^+ + O + e^-\\
CO_2 + e* \longrightarrow CO + O
\end{eqnarray}
to enable model experiments in $CO_2$-rich atmospheres.
\subsection{GARLIC}\label{subsec:GARLIC}
The \textit{Generic  Atmospheric  Radiation  Line-by-line  Infrared  Code} \citep[\textit{\textit{GARLIC}}, e.g.][]{schreier2014, schreier2018agk, schreier2018ace} is used for spectral analysis utilizing the atmospheric profiles (in particular p, T, and the composition) derived within our model suite.
\subsection{Model Coupling}
This study focuses on the impact of cosmic rays on atmospheric ionization and ion chemistry. Therefore, we slightly modified the INCREASE model suite discussed in \citet{Herbst-etal-2019c}. As shown in Fig.~\ref{fig:1}, to derive the results presented in Section~\ref{sec:results}, the following steps are performed: (1) measured stellar UV fluxes are used as input for \textit{1D-TERRA} and \textit{ExoTIC}, (2) the incoming cosmic ray fluxes at the location of TRAPPIST-1e are derived either by analytic or numerical approaches (see Sec.~\ref{sec:GCRs} and \ref{sec:SEPspectra}), (3) \textit{1D-TERRA} provides vertical profiles of temperature, pressure, and trace gases to \textit{ExoTIC} and \textit{AtRIS}, (4) to account for a potential planetary magnetic field, \textit{PLANETOCOSMICS} is utilized to provide top-of-the-atmosphere (TOA) particle fluxes as input for the computations performed with \textit{AtRIS}, (4) utilizing \textit{AtRIS}, the GCR and StEP-induced altitude-dependent radiation exposure and atmospheric ionization down to the exoplanetary surface are modeled ; the latter is further used as input by \textit{ExoTIC}, (5) with \textit{ExoTIC}, the impact of changing atmospheric ionization for the different atmospheric compositions and parameterization of the neutral atmosphere is determined, (6) the changes in neutral-ion chemistry are computed with \textit{ExoTIC}, (7) the resulting global atmospheric composition and temperature vertical profiles are utilized to compute atmospheric transit (primary) spectra with \textit{GARLIC}.
\section{Simulation Setup}
\subsection{Stellar Parameters and TRAPPIST-1 Spectra}

In this study we assumed a stellar effective temperature of $T_{eff}$ = 2516 K \citep{VanGrootelEA2018}, a stellar radius of $R_{\star}$ = 0.124 R$_{\sun}$ \citep{Kane-2018}, a stellar mass of $M_{\star}$ = 0.089 $M_\sun$ \citep{VanGrootelEA2018}, and a distance of 12.43 pc \citep{Kane-2018}. Since the stellar spectral energy distribution in the (E)UV can strongly impact the photochemistry of terrestrial exoplanetary atmospheres \citep[e.g.,][]{GrenfellEA2013, TianEA2014} we use the semiempirical TRAPPIST-1 model spectrum by \citet{WilsonEA2021} based on observational data from XMM-Newton (X-ray regime) and from the Hubble Space Telescope (HST; in the 13 – 570 nm range) with a gap between 208 and 279 nm obtained through the Mega-MUSCLES Treasury survey \citep[e.g.,][]{FroningEA2018, WilsonEA2019}. Further, we use data from \citet{WilsonEA2021} who filled the higher wavelength range based on a PHOENIX photospheric model \citep{BaraffeEA2015, AllardEA2016}.

\subsection{Scenarios for Planetary Atmospheres}\label{sec:Scenarios}
TRAPPIST-1e has a radius of 0.92 $R_\earth$, a mass of 0.692 $M_\earth$, and orbits TRAPPIST-1 within 6.1 days at a distance of 0.02925 AU\footnote{\url{https://exoplanets.nasa.gov/exoplanet-catalog/3453/TRAPPIST-1-e/}}. \citet{Wunderlich-etal-2020} investigated multiple atmospheric scenarios for the TRAPPIST-1 planets e and f. As a follow-up to their study, we utilize six of their discussed atmospheric compositions, adding the impact of cosmic rays to the picture to investigate their impact on the scenario-dependent atmospheric biosignatures. The following scenarios are assumed: 
\begin{itemize}
\itemsep0em 
    \item[$\circ$] \textbf{Dry-dead} atmospheres, without an ocean and only taking into account volcanic outgassing, and thus assuming a relative humidity of 1\% with 0.1 bar of CO$_2$ \textbf{[1] (black solid lines}) and 1 bar of CO$_2$ \textbf{[2](black dashed lines}).
    \item[$\circ$] \textbf{Wet-dead} atmospheres, with an ocean and only taking into account volcanic outgassing, and thus assuming a relative humidity of 80\% without biosphere emissions and with 0.1 bar of CO$_2$ \textbf{[3] (blue solid lines}) and 1 bar of CO$_2$ \textbf{[4] (blue dashed lines}).
    \item[$\circ$] \textbf{Wet-alive} atmospheres, with an ocean and terrestrial biogenic and volcanic fluxes, similar to scenarios \textbf{[3]} and \textbf{[4]} with 0.1 bar of CO$_2$ \textbf{[5] (green solid lines}) and 1 bar of CO$_2$ \textbf{[6]  ( green dashed lines}), and high levels of oxygen from biogenic emissions.
\end{itemize}
%
\begin{table*}[!t]
\centering
    \begin{tabular}{c c|c|c|c|c|c|c|c}
       \hline
        & & $P_{\mathrm{surface}}$ [hPa] & CO$_2$ [\%] & N$_2$ [\%] & Ar [\%] & H$_2$O [\%] & O$_2$ [\%] & others [\%]\\
        \hline
                       & dry-dead [1] & 1031.62 & 13.3 & 84.0 & 1.3 & 3.47$\times 10^{-4}$ & 2.14$\times 10^{-1}$ & 1.19 
\\
        0.1 bar CO$_2$ & wet-dead [3] & 1035.02 & 13.4 & 85.0 & 1.2 & 1.49$\times 10^{-1}$ & 7.01$\times 10^{-3}$ & 0.24
\\
                       & wet-alive  [5] & 1036.45 & 12.8& 51.7 & 1.2 & 1.95$\times 10^{-1}$& 34.04 & 0.065
\\
        \hline
        \hline
                       & dry-dead  [2] & 1879.76 & 61.1 & 34.4 & 1.0 & 2.24$\times 10^{-3}$ & 1.14 & 2.36
\\
        1 bar CO$_2$   & wet-dead [4] & 2043.32 & 63.0 & 32.0 & 1.0 & 0.3 & 4.27$\times 10^{-3}$ & 3.69\\
                       & wet-alive [6] & 2079.23 & 61.2 & 1.13 & 1.0& 3.82 & 32.77 & 0.08\\
        \hline
    \end{tabular}
    \caption{Scenario-dependent mass fractions of atmospheric constituents that have been utilized as input for \textit{AtRIS}, \textit{1D-TERRA}, and \textit{ExoTIC}. Scenario numbers are shown in square brackets.}
    \label{tab:1}
\end{table*}
\begin{figure}[!t]
\vspace{0.3cm}
    \centering
    \includegraphics[width=0.7\columnwidth]{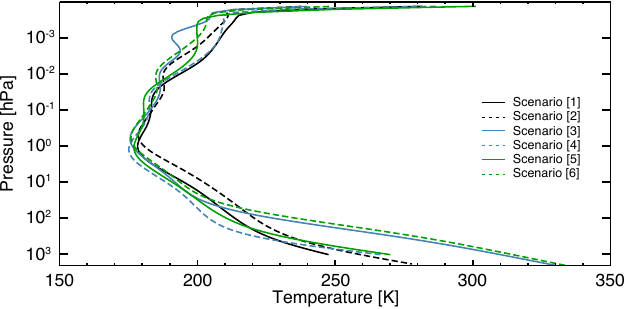}
    \caption{Temperature profile of the six scenarios discussed in \citet{Wunderlich-etal-2020} (see Sec.~\ref{sec:Scenarios}, colored solid lines) in comparison to the terrestrial profile (dashed blue line).}
    \label{fig:4}
\end{figure}
Figure~\ref{fig:4} shows the temperature profile of all investigated scenarios (colored solid lines) while Tab.~\ref{tab:1}  gives the surface pressure and scenario-dependent mass fractions of atmospheric CO$_2$, N$_2$, Ar, H$_2$O, and O$_2$ at the planetary surface that have been utilized as input for \textit{1D-TERRA}, \textit{AtRIS}, and \textit{ExoTIC}. 

\subsection{Modeling the Ion-Chemistry Changes due to a Carrington-Like Event}
This study assumes TRAPPIST-1e to be a tidally-locked planet and with a stellar zenith angle fixed to $60^\circ$ to mimic the global mean chemistry. Time-dependent experiments to model the impact of the Carrington-like event (see Sec.~\ref{sec:SEPspectra}) for all six scenarios were carried out for 216 Earth hours, during which the stellar irradiance was considered constant. To model the scenario-dependent ion-chemistry changes induced by the event, we implemented the corresponding ion pair production rate profiles provided by \textit{AtRIS} (see results discussed in Sec.~\ref{sec:IPPR_res}) for a time of six Earth hours after a spin-up period of 24 hours. As a first-order approximation, throughout these six hours, the stellar particle flux is assumed to be constant and zero otherwise, a simplification often used in solar-terrestrial studies to assess longer-lasting impacts of solar particle events \citep[e.g.,][]{Jackman2000, Jackman2005, Jackman2011, Matthes2017}. In addition, a model run without energetic particle impact and the particle event but identical in every other aspect was performed as a reference for each scenario.

\section{Results}\label{sec:results}
The following investigates the CR-induced changes in atmospheric ion-pair production, ion chemistry, and composition.
\subsection{The Cosmic Ray Induced Atmospheric Ion Pair Production Rate Changes}\label{sec:IPPR_res}

\begin{figure}[!t]
    \centering
    \includegraphics[width=0.7\columnwidth]{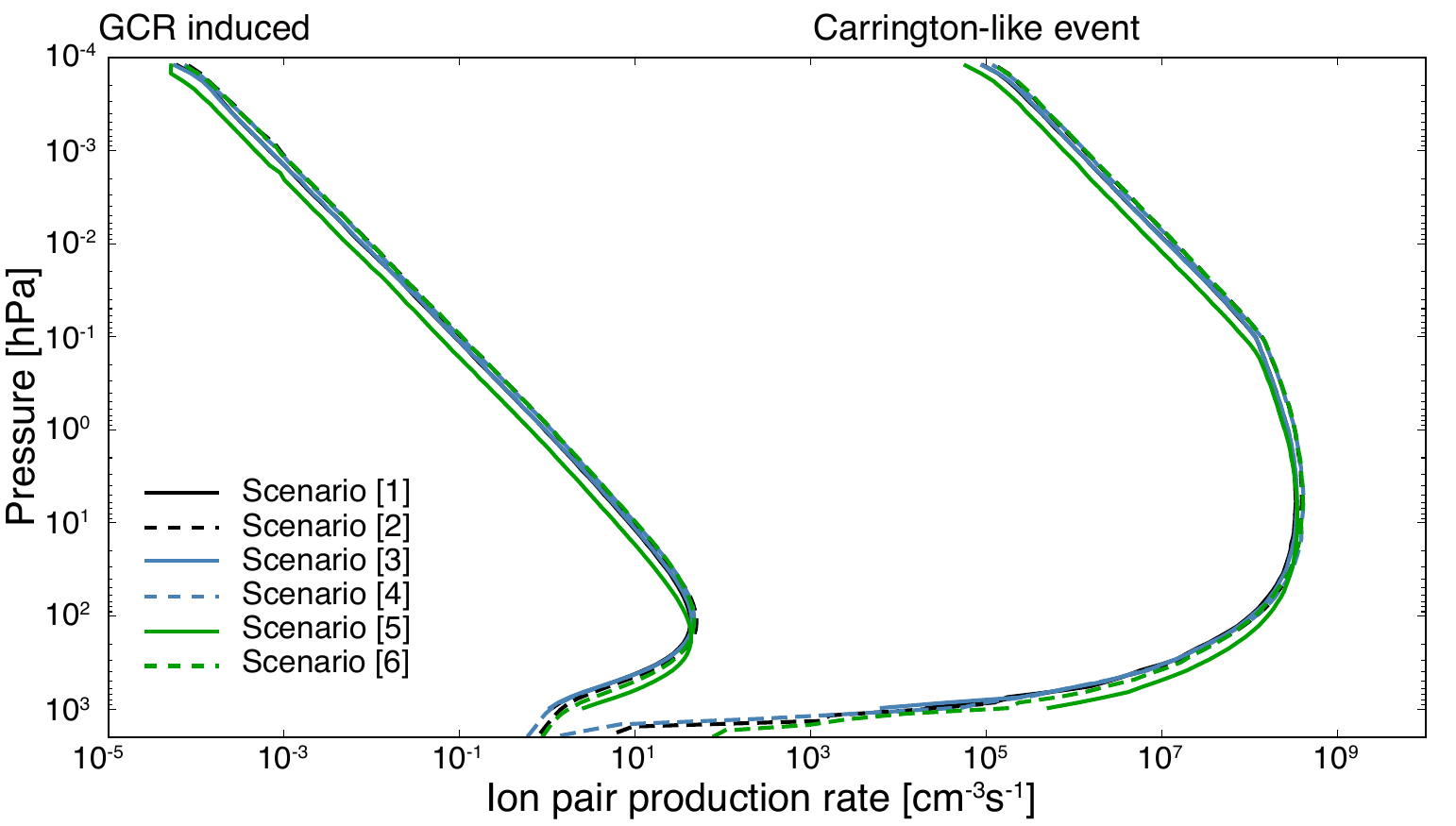}
    \caption{Atmospheric ion pair production rates for scenarios [1] to [6] (colored lines) caused by GCRs (left side of the figure) and the Carrington-like event discussed in \citet{Vida-etal-2017} (right side of the figure). In addition, the terrestrial production rate values during similar conditions are shown as blue dashed lines.}
    \label{fig:6}
\end{figure}

Figure~\ref{fig:6} shows the modeled scenario-dependent CR-induced ion-pair production rate values.  As can be seen, the altitude of the GCR-induced production rate maximum varies from 200 -- 100~hPa (12-15 km) (wet-alive 0.1 bar CO$_2$, green solid line) up to 100 -- 50~hPa (15-22 km) (wet-dead 1 bar CO$_2$, dashed green line). Further, the production rate maximum values caused by the Carrington-like event are about eight orders of magnitude higher than those during the quiet stellar conditions, and significant differences at the surface - depending on the atmospheric scenarios - occur. It shows that the higher the pressure (i.e., the CO$_2$ content), the lower the surface production rates due to CR shielding by CO$_2$. In particular, the Carrington-like event for the wet-dead 1 bar CO$_2$ scenario (in dashed blue) results in lower production rate values than the GCR-induced wet-alive 0.1 bar CO$_2$ scenario (in green). 
%
\subsection{The Cosmic Ray Induced Atmospheric Radiation Exposure}\label{sec:radiation}

\begin{figure}
    \centering
    \includegraphics[width=0.7\columnwidth]{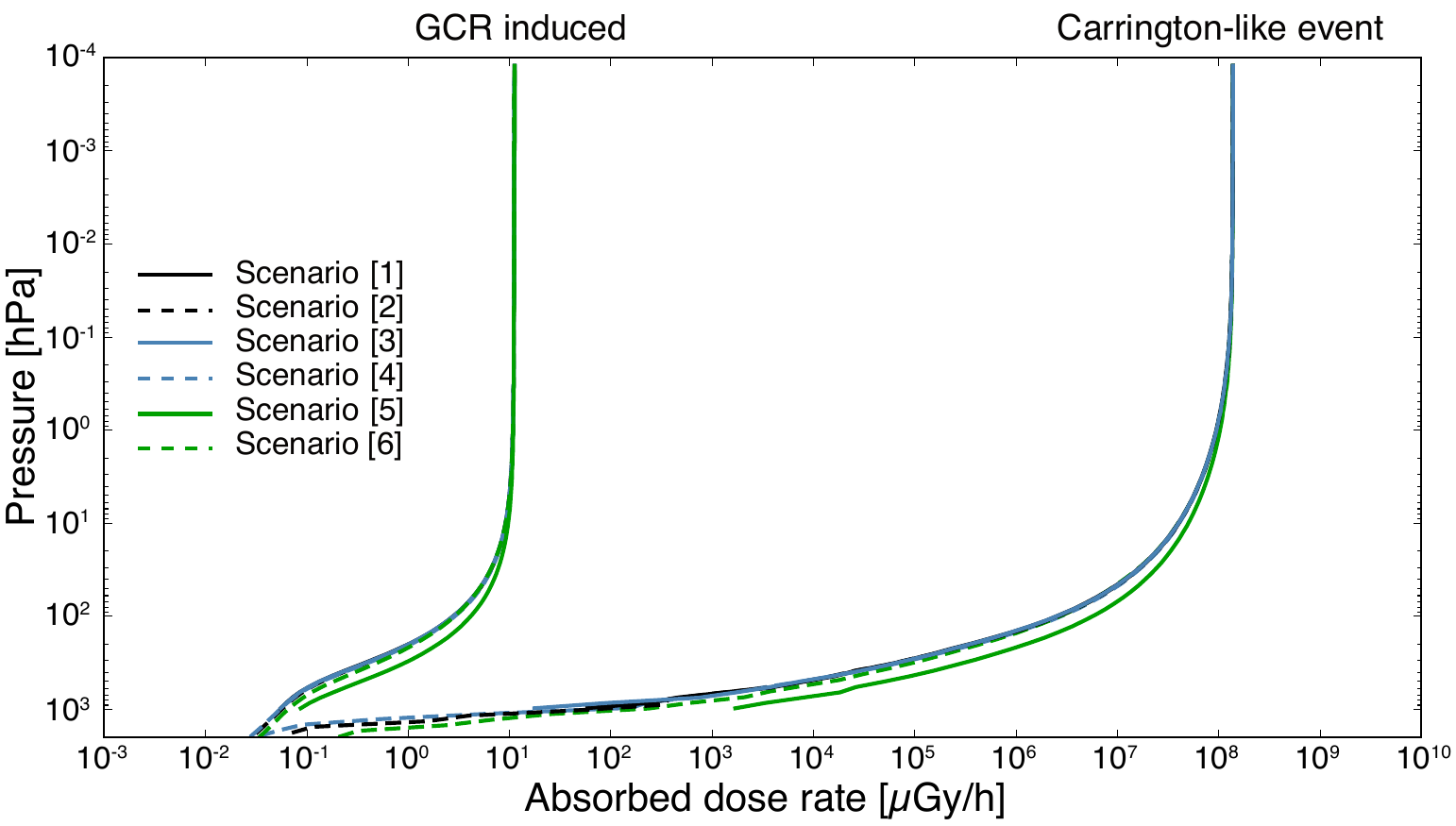}
    \caption{Averaged atmospheric absorbed dose rates for scenarios [1] to [6] (colored lines) caused by GCRs (left side of the figure) and the Carrington-like event discussed in \citet{Vida-etal-2017} (right side of the figure). In addition, the terrestrial absorbed dose rates during similar conditions are shown as blue dashed lines.}
    \label{fig:7}
\end{figure}

The resulting altitude-dependent absorbed dose rates of the six scenarios are shown in Fig.~\ref{fig:7}. Here, profiles without energetic particle impact (left side of the figure) and caused by the Carrington-like event (right side) are displayed. As can be seen, all scenarios are similar within $\pm$ 2\% at altitudes above 10~hPa (40 km) (1~hPa (55 km)) during quiescent stellar conditions (during the Carrington-like event). 

In the CO$_2$-rich scenarios  [2,4,6], a Carrington-like event only leads to comparatively negligible radiation exposure increases at the planet's surface. However, in the CO$_2$-poor scenarios [1,3,5], particularly within the wet-alive atmosphere, a Carrington-like event would lead to more than three to four orders of magnitude increase at the planetary surface. As listed in Tab.~\ref{tab:3}, in the CO$_2$-poor wet-alive case, the Carrington-like event would lead to absorbed dose rates around 1.69 mGy/h, which is more than 23,300 times higher than those during the actual Carrington event at the terrestrial surface. To put this into perspective, note that chromosomal aberrations and mutations can already occur at dose rates between 1 mGy/h and 20 mGy/h, which corresponds to accumulated doses of 0.5 to 8 Gy \citep[e.g., ][and references therein]{Ruehm-etal-2018}. Thus, although a Carrington-like event in a CO$_2$-poor wet-alive atmosphere would lead to changes at DNA level, it would not be lethal.
\begin{table}[!t]
    \centering
    \begin{tabular}{c|c|c|c}
    \hline
    &Scenario & $D_{\mathrm{GCR}}$ [$\mu$Gy/h] & $D_{\mathrm{SEP}}$ [$\mu$Gy/h] \\
    \hline
     &[1] & 5.54$\times 10^{-2}$ & 5.85$\times 10^{1}$\\
    0.1 bar CO$_2$ & [3] & 5.51$\times 10^{-2}$ & 1.76$\times 10^{1}$\\
     &[5] & 8.68$\times 10^{-2}$ & 1.69$\times 10^{3}$\\
     \hline
     \hline
     &[2] & 3.24$\times 10^{-2}$ & 7.42$\times 10^{-2}$\\
     1 bar CO$_2$ & [4] & 2.85$\times 10^{-2}$ & 3.54$\times 10^{-2}$\\
     & [6] & 3.45$\times 10^{-2}$ & 2.14$\times 10^{-1}$ \\
     \hline      
    \end{tabular}
    \caption{Absorbed dose rate values at the surface of TRAPPIST-1e for the Earth-like atmosphere scenarios [1] to [6] during stellar quiet conditions ($D_{\mathrm{GCR}}$) and the Carrington-like event ($D_{\mathrm{SEP}}$) scaled after \citet{Vida-etal-2017}.  }
    \label{tab:3}
\end{table}
\subsection{Atmospheric Ion Chemistry and Composition Changes}

The impact of the Carrington-like event on the atmospheric chemical composition of scenarios [1] - [6] is shown in the panels of Fig.~\ref{fig:14}, displaying the difference between the mean volume mixing ratios [ppm/ppb] with and without ion chemistry during the event, for which an event period of 6 hours is assumed. This allows us to visualize the production and destruction of individual species due to the impact of atmospheric ionization and subsequent ion and neutral chemistry. 

\subsubsection{Water Loss and HOx Formation}
\begin{figure}
\centering
    \includegraphics[width=0.65\linewidth]{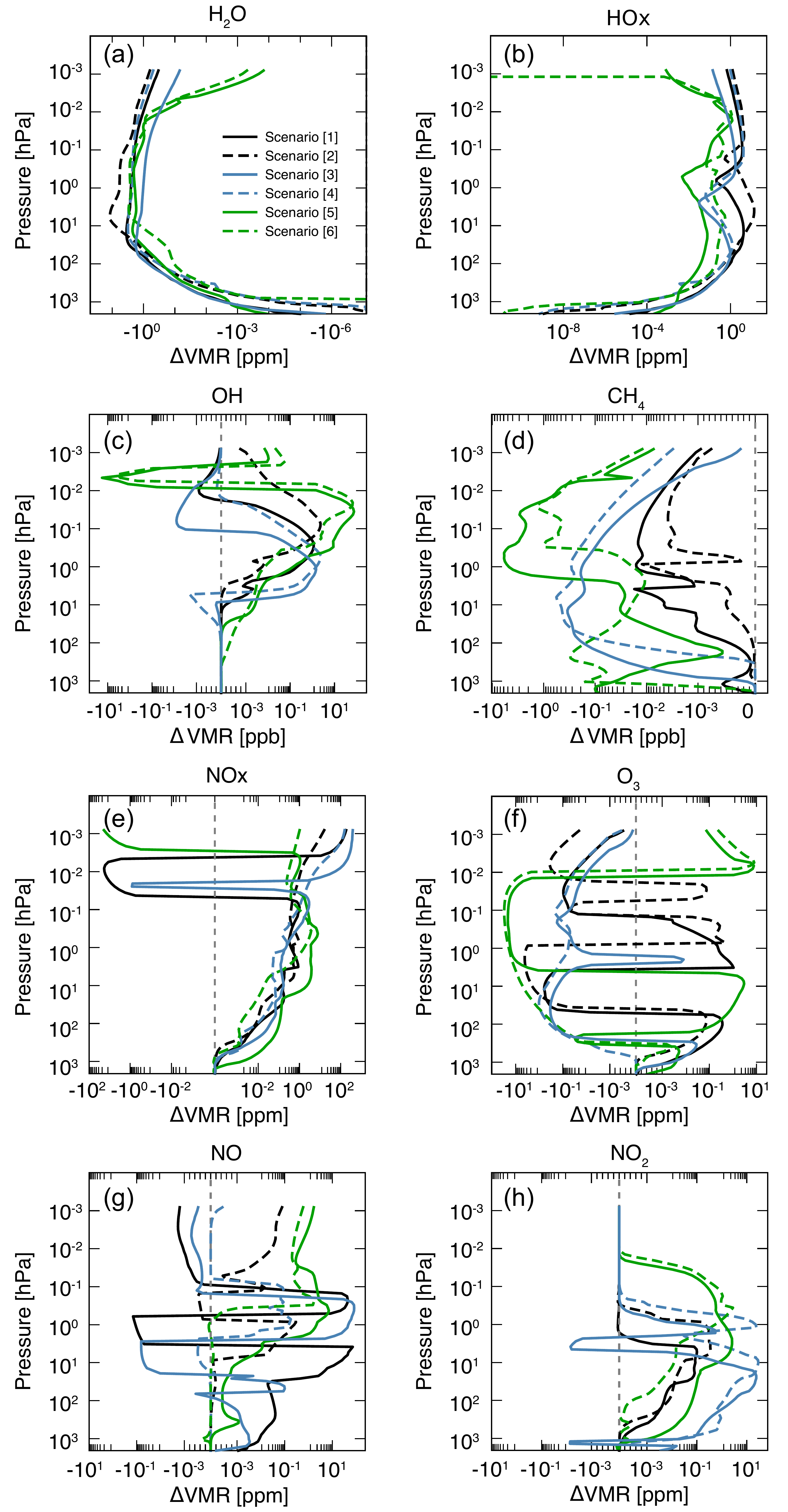}
    \caption{Absolute differences between model experiments with and without atmospheric ionization for the time of a six-hour Carrington-like stellar particle event modeled with \textit{ExoTIC}.}
     \label{fig:14}
\end{figure}

As shown in Fig.~\ref{fig:14}, results suggest the destruction of water vapor (H$_2$O, panel (a)) throughout the entire atmosphere, which can be explained by the uptake of water vapor in positive water cluster ions that ultimately leads to the formation of HOx during recombination of the water cluster ions \citep{Sinnhuber-etal-2012}. The loss of water vapor is mirrored by an increase of HOx of comparable size as shown in panel (b). The relatively high water loss values in the dry-dead scenarios [1,2] result from greater amounts of water vapor in the middle atmospheres above 200~hPa (10~km).
\subsubsection{Methane Change}
At around 10 -- 0.01~hPa (40 -- 70~km), the HOx formed by the ion chemistry is mainly in the form of hydroxyl (OH, panel (c)). Hydroxyl acts as a sink for methane (CH$_4$, panel (d)) via the reaction CH$_4$ + OH, and a decrease of methane over nearly the whole model altitude is evident, maximizing in the altitudes where the OH increase is most prominent in the respective scenario. The largest increases in OH, and respective decreases in methane, are observed in the wet-alive scenarios [5-6] between 0.1 -- 0.01~hPa (50 -- 70~km) due to more available O$_2$ and H$_2$O.

\subsubsection{Ozone Change}
Based on the strong increase in HOx from 0.01--10~ppm at 200 -- 0.01~hPa (10 -- 70~km) in all scenarios, and the increase of NOx of a few ppb to several ppm depending on scenario and altitude (see panels (e), (g) and (h) of Fig.~\ref{fig:14}), a loss of ozone would be expected over a wide range of altitudes for all scenarios. The change in ozone for all six scenarios is shown in panel (f). As can be seen, ozone loss is observed in all scenarios. However, the altitude range differs considerably between the various scenarios, with different regions of ozone formation in some altitudes and scenarios. For example, for the CO$_2$-rich wet-dead scenario [4], ozone loss extends nearly over the whole altitude range, from about 200~hPa (10~km) to the top of the model boundary above 0.01~hPa (80~km). In contrast, for the CO$_2$-poor wet-alive scenario [5], ozone loss is also observed in 10 -- 0.01~hPa (40 -- 70~km), while above and below this altitude range, ozone formation is observed. The CO$_2$-rich wet-alive scenario [6] shows a similar behavior, with, however, a larger area of ozone loss than the CO$_2$-poor scenario between 200 -- 0.01~hPa (15 -- 70~km), but similar regions of ozone formation above and below. In the dry-dead scenarios [1,2] and in the CO$_2$-poor wet-dead [3] scenario, ozone loss and ozone formation regions alternate between 200 hPa (10~km) and 0.01~hPa (70~km), with ozone loss dominating at the top. Ozone formation at altitudes above 0.01~hPa (70~km) is likely due to the dissociation of O$_2$ by charged particle impact ionization with subsequent ozone formation via O+O$_2$ in the oxygen-rich atmospheres of the wet-alive scenarios [5,6] at altitudes where water vapor is not abundant, and HOx formation therefore small (see Fig.~\ref{fig:14} (b)). Another mechanism of ozone formation that could be initiated by atmospheric particle impact ionization is the so-called \textit{smog}-mechanism, where the photodissociation of NO$_2$ in the visible spectral range provides atomic oxygen (O) for ozone formation. Similar reaction chains lead to ozone formation, especially in the polluted summer troposphere of Earth but not in the terrestrial stratosphere and mesosphere, where the availability of atomic oxygen from photolysis of O$_2$ in the far UV spectral range enables catalytic ozone loss involving NOx \citep[e.g.,][]{Lary1997}. The M-dwarf TRAPPIST-1 has a much redder spectrum than the Sun, with significantly lower radiative fluxes in the UV spectral range. Therefore, photodissociation of O$_2$ is negligible, and the formation of NOx by particle impact ionization leads to ozone formation via the smog mechanism rather than catalytic ozone loss as in the terrestrial stratosphere. Catalytic ozone loss via HOx does not depend on the availability of atomic oxygen \citep{Lary1997}, and can also occur in our scenarios. The distinction between ozone loss and ozone formation, therefore, depends critically on the availability of NOx, HOx, and - only in the upper part of the atmosphere - O$_2$, leading to the layering of ozone formation and loss calculated, e.g., in the dry-dead scenarios [1,2], and to the distinctive difference between the CO$_2$-rich [6] and CO$_2$-poor [5] wet-alive scenarios, which differ in the amount of HOx production above 100~hPa (15~km).

\begin{figure}[!t]
\vspace{0.3cm}
    \centering
    \includegraphics[width=0.7\linewidth]{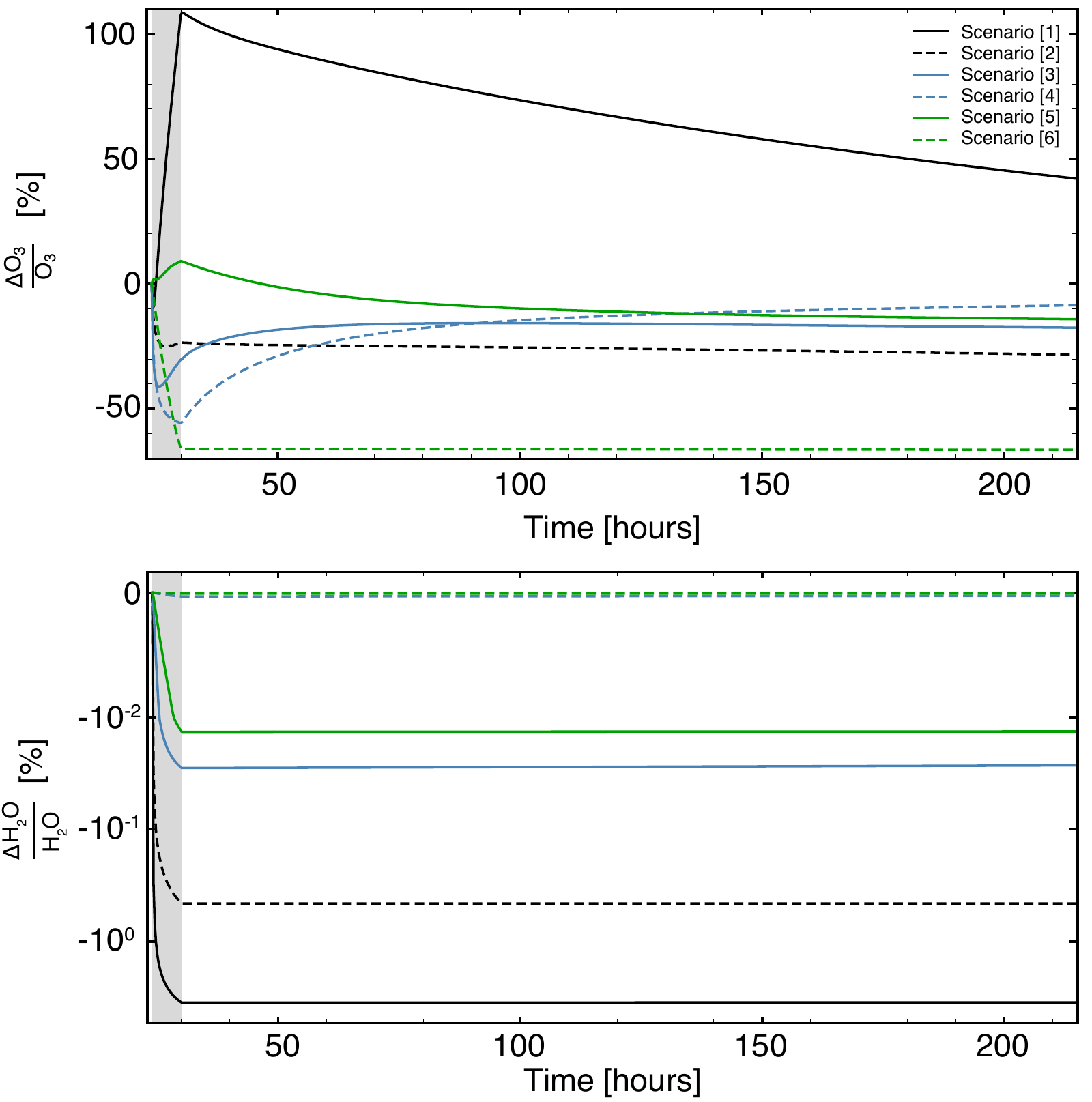}
    \caption{Time series of the deviation in the O$_3$ column density (upper panel) and the H$_2$O column density (lower panel)} for the six scenarios during (grey bar) and after the particle event.
    \label{fig:12}
\end{figure}
The upper panel of Fig.~\ref{fig:12} shows the total ozone column density, reflecting the consequences of ozone loss versus ozone formation: At the start of the event, ozone loss becomes apparent in most scenarios. However, the CO$_2$-poor dry-dead [1] and wet-alive [5] scenarios differ from the other scenarios and show an increase in total column ozone. These scenarios show particularly strong ozone increases around 100 -- 10~hPa (15 -- 40~km) (see also Fig.~\ref{fig:14}(f)) at the lower edge of the ozone layers in these particular atmospheres where relative changes in ozone have the largest impact on the total column amount. This is also indicated in the CO$_2$-rich dry-dead scenario [2]. However, ozone formation around 90~hPa (20~km) is smaller and, therefore, does not lead to production in the total ozone column. 

The evolution of ozone after the particle event is determined by a sensitive balance of production and loss mechanism involving the amount of O, HOx, and NOx available at the end of the event as well as the solar irradiance spectrum. Relevant mechanisms include ozone loss by the short-lived HOx as well as ozone formation by the much longer-lived NOx in the lower part of the atmosphere, by atomic oxygen in the upper part of the atmosphere, and by UV photolysis of O$_2$ and CO$_2$ in the upper part of the atmosphere. Ozone loss and formation by HOx and NOx can further be modulated by the formation of HNO$_3$ via the reaction OH+NO$_2$, which reduces both HOx and NOx. Consequently, the evolution of ozone after the event is very different in the different scenarios. After the event, ozone columns recover to a value similar to the starting point over the course of a few days in the wet-dead scenarios [3,4] and the CO$_2$ poor wet-alive scenarios [5]. In scenario [1], total ozone recovers from strongly enhanced values much slower and is still greatly enhanced at the end of the model period, while in the CO$_2$-rich dry-dead [2] and wet-alive [6] scenarios,  ozone stays greatly depleted without any recovery throughout the model period. In these two scenarios, the formation of NOx in the lower part of the atmosphere where the NOx-driven ozone formation dominates is comparatively low due to the lower amount of N$_2$ available in the CO$_2$ rich scenarios, while the amount of HOx produced is high. Additionally, the lower panel of Fig. 7 shows the relative changes of the H$_2$O column density. As in Fig.~\ref{fig:14}(a), water loss of up to 2\% during the event can be observed, especially for the dry and dead scenarios [1,2]. However, the column density calculation is dominated by the lower parts of the atmosphere, so the percentual change of the total water column is small (see Table~\ref{tab:1}). This is especially true for the wet scenarios [3--6]. The water content does not recover within the model experiment period, as HOx preferentially forms $H_2$, not water vapor.

Thus, the impact of a Carrington-like event and the corresponding particle impact ionization on the transmission spectra can likely be long-lasting in some scenarios and might be considered particularly in the surroundings of a very active star with frequent flares.
The varying response of the total amount of ozone during and after the particle event 
will have implications for the transmission spectra and limit the usability of ozone as a biosignature in planets orbiting active cool stars if the atmospheric composition is not very well characterized. 
\begin{figure*}[!t]
    \centering
    \includegraphics[width=\linewidth]{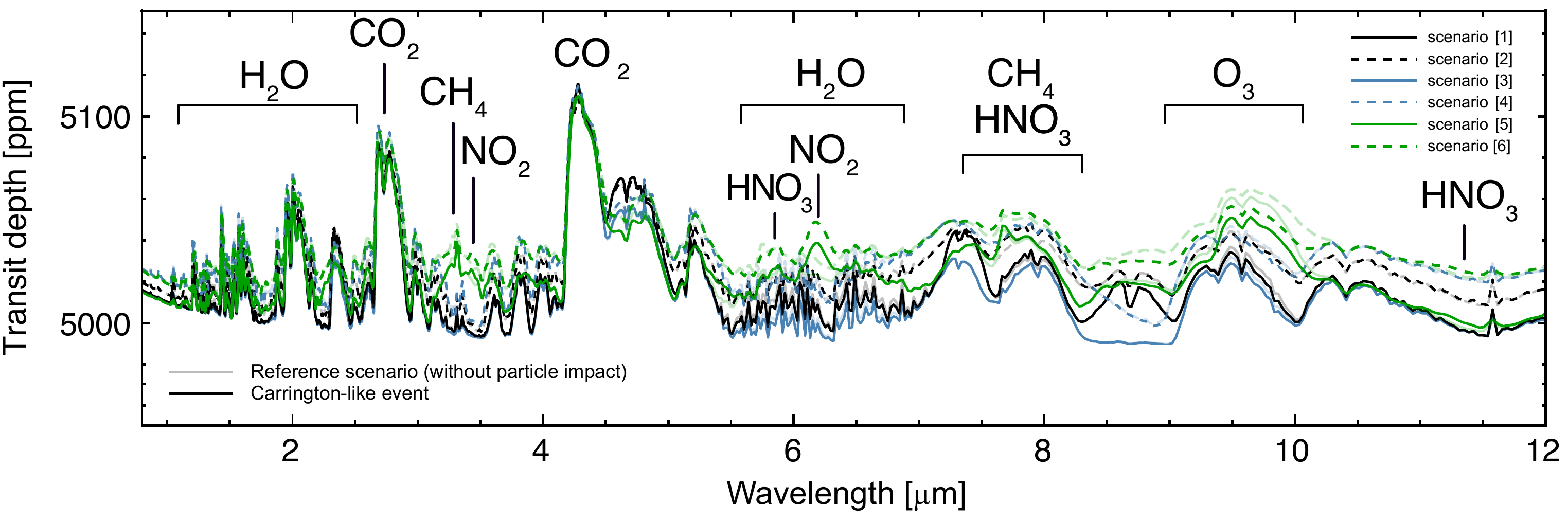}
    \caption{Modeled scenario-dependent transit depths of TRAPPIST-1e in parts per million (ppm) over wavelength $\lambda$ (R=300), showing the spectra without energetic particle impact (solid lines) and during the Carrington-like event (dashed lines).}
    \label{fig:TDcomp}
\end{figure*}
\subsection{Consequences for the Transmission Spectra}\label{sec:biosignatures}
Figure~\ref{fig:TDcomp} shows the transmission spectra of the six scenarios generated with GARLIC \citep{schreier2014, schreier2018agk, schreier2018ace} neglecting the impact of energetic particles (reference case, weak lines) and the Carrington-like event (strong line). Thereby, the time mean over the 6-hour event period was used to calculate the mean atmospheric state. GARLIC used a resolving power of \mbox{R = 300} to generate the transmission spectra \citep[see,][]{Wunderlich-etal-2020}. As can be seen, the Carrington-like event - highlighting the impact of the enhanced atmospheric ionization on the transmission spectra - leads to significant changes in the scenario-dependent transmission spectra. 

\begin{figure}[!t]
    \centering
        \centering
        \includegraphics[width=0.7\columnwidth]{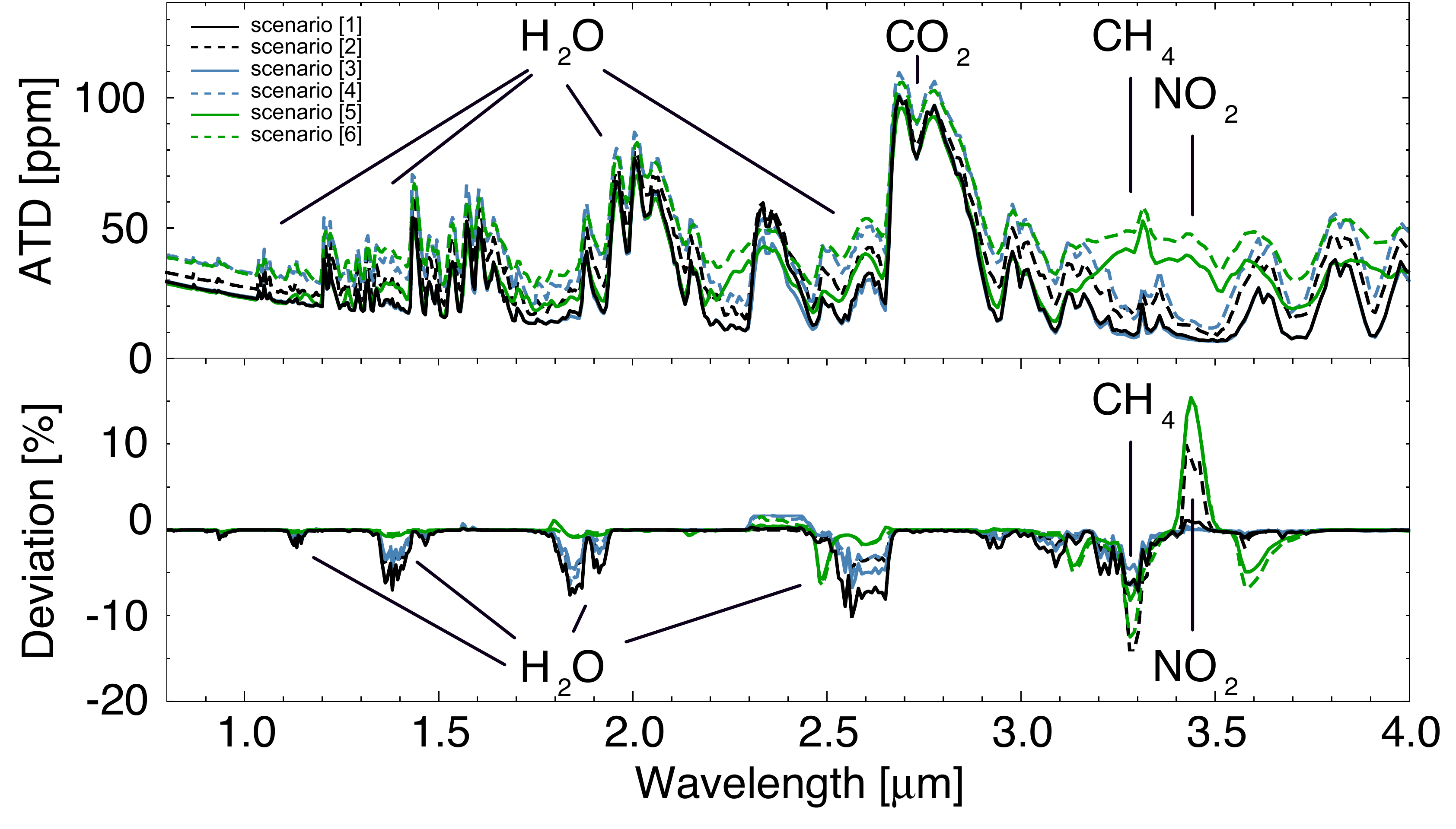}
        \includegraphics[width=0.7\columnwidth]{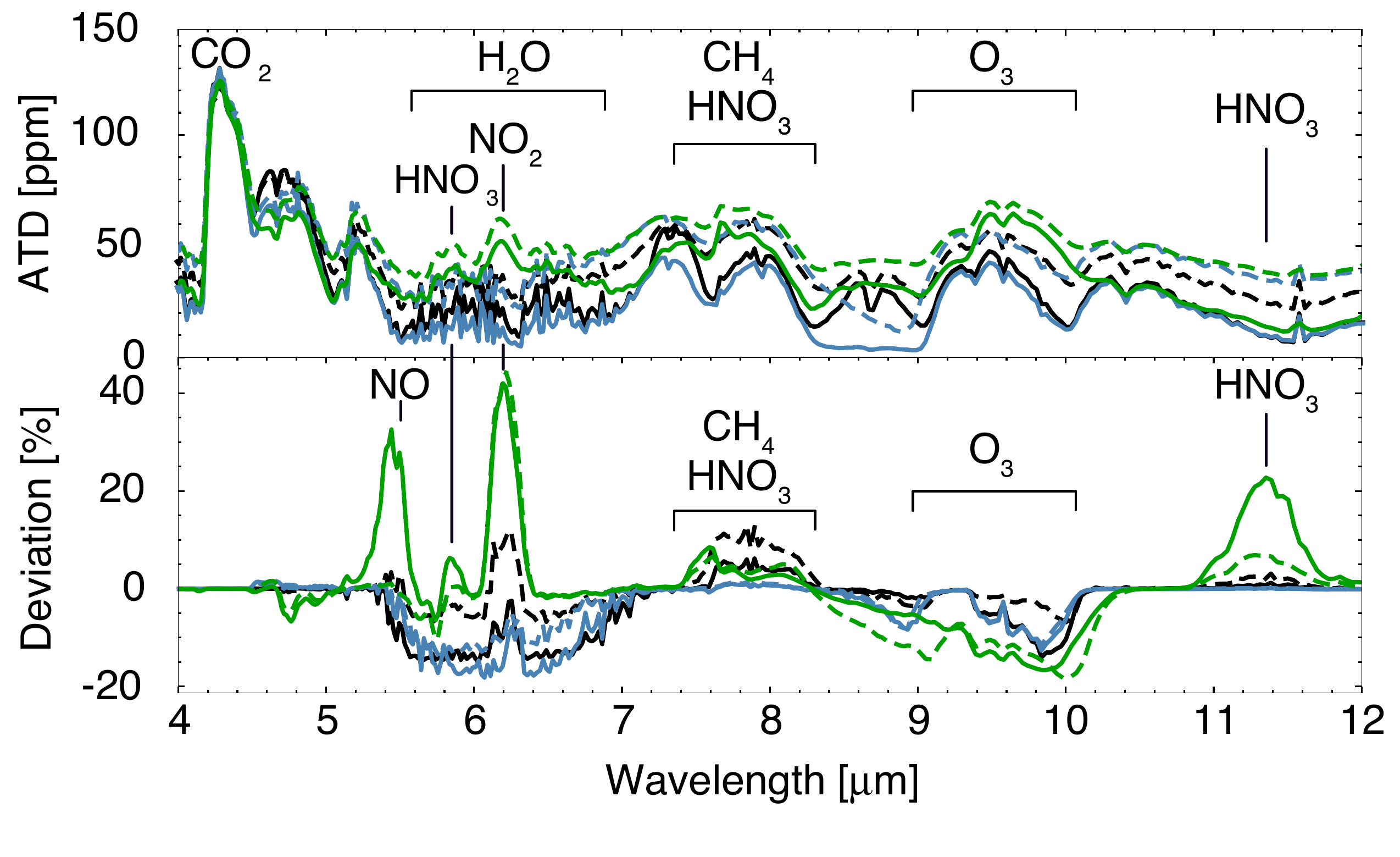}
    \caption{Resulting transmission spectra of the six scenarios discussed in Section~\ref{sec:Scenarios} based on the mean atmospheric composition during the 6-hour Carrington-like event (Panels 1 and 3 in Atmospheric Transit Depth (ATD)). The second and fourth panels show the deviation (with - without ions) in each transmission spectra compared to a corresponding reference run without ion chemistry.
    }
    \label{fig:13}
\end{figure}

For a better visibility, Fig.~\ref{fig:13} shows the spectra during the Carrington-like event in the first (0.8 -- 4.0 $\mu$m) and third panel (4.0 -- 12.0 $\mu$m). In contrast, the differences between the respective spectra and the reference case neglecting the impact of energetic particles are displayed in the second and fourth panels, respectively. 

\begin{figure}[!t]
    \centering
    \includegraphics[width=0.7\columnwidth]{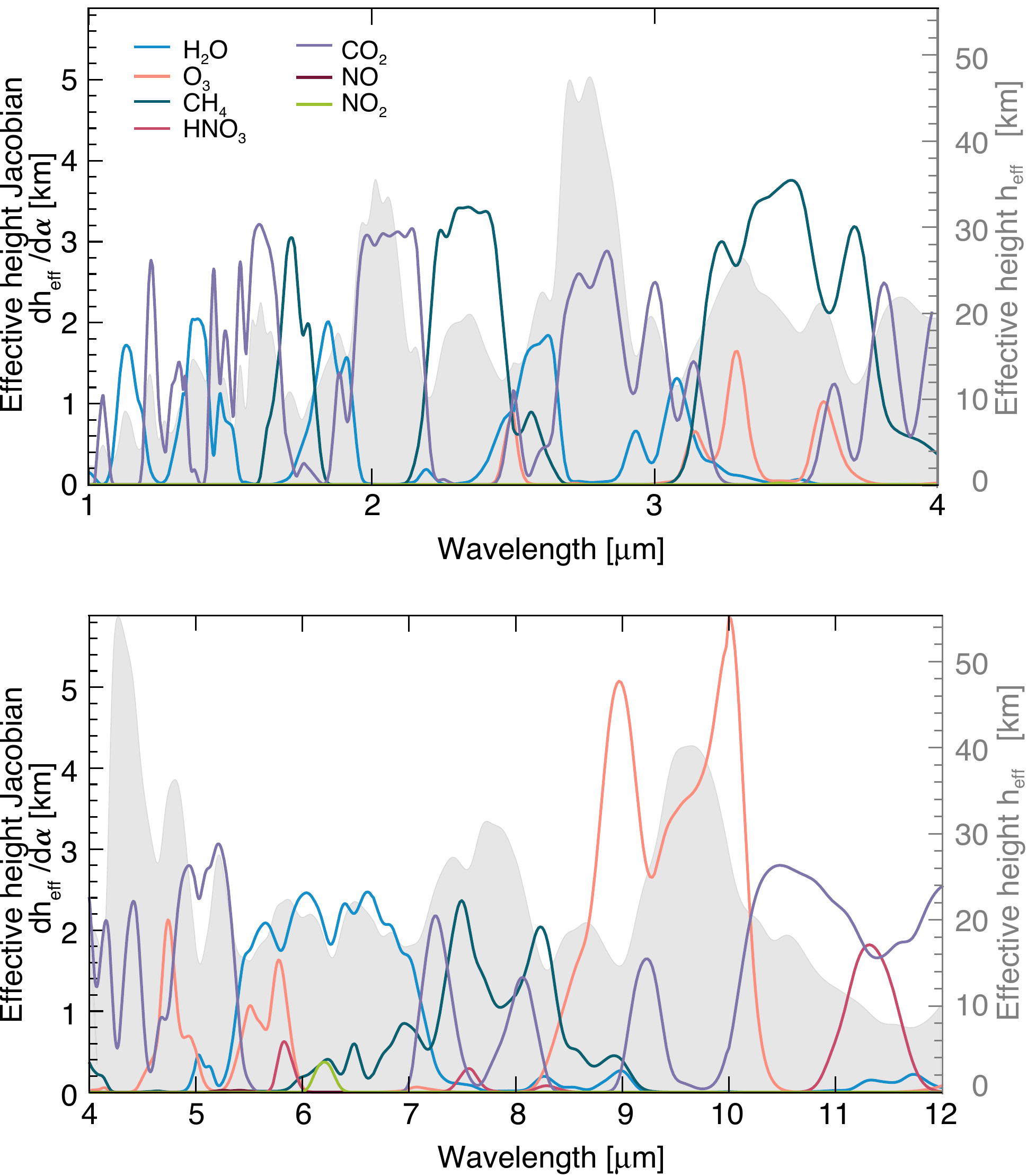}
    \caption{Jacobian of the effective height spectra of the trace gases shown in Fig.~\ref{fig:13} (R = 100). Grey areas: Effective height - the altitude where the atmosphere becomes optically dense ($\tau > 1$) - indicating that in the 4 -- 12~$\mu$m range, the spectral signatures all derive from the middle atmosphere above 9~km. Colored lines: the derivatives of the effective heights to the most important species, a measure of the contribution of the trace gases to the spectral signal. Shown as an example is the CO$_2$-poor wet and alive scenario ([5]).}
   \label{fig:franz}
\end{figure}

To help identity which trace gases contribute to the spectral signatures identifiable in Fig.~\ref{fig:13}, the Jacobian, i.e., the partial derivatives of the effective height w.r.t. to the scale factor $\alpha$ of the molecular density, of the effective height spectra regarding the trace gases are shown in Fig.~\ref{fig:franz} \citep[colored lines, here for $\alpha$ = 1, see][subsection 3.3 for further information]{Schreier15}. In the spectral range 0.7--4~$\mu$m, overlapping signatures of CO$_2$ (1 -- 1.1~$\mu$m, 1.2 -- 1.4~$\mu$m, 1.4 -- 1.7~$\mu$m,  1.8 -- 2.3~$\mu$m, 2.4 -- 3.2~$\mu$m, 3.5 -- 4~$\mu$m), CH$_4$ (1.6 -- 1.83~$\mu$m, 2.1 -- 2.7~$\mu$m, 3.0 -- 4.0~$\mu$m) and H$_2$O (1.1 -- 1.25~$\mu$m, 1.25--1.6~$\mu$m, 1.7--2.0~$\mu$m, 2.1--2.7~$\mu$m, 2.8--3.4~$\mu$m) dominate, with smaller contributions of O$_3$ (3--3.8~$\mu$m) and NO$_2$ (3.4--3.5~$\mu$m) masked by the larger signals.

In the range 4--12~$\mu$m, spectral signatures of CO$_2$ (4--5.5~$\mu$m, 7--8.3~$\mu$m, 8.9--12~$\mu$m), CH$_4$ (5.8--9.2~$\mu$m), H$_2$O (4.9--7.7~$\mu$m) and O$_3$ (8.1--10.4~$\mu$m) dominate, with smaller contributions from HNO$_3$ (4.7--6~$\mu$m, 7.3--7.8~$\mu$m, 8.1--8.5~$\mu$m, 8.8--12~$\mu$m), NO$_2$ (6--6.4~$\mu$m) and NO (5--5.7~$\mu$m). Thus, for example, the strong water signal at 2.1 -- 2.7~$\mu$m overlaps with the CO$_2$ band at 2.4--3.2~$\mu$m which is weaker in the maximum of the H$_2$O band around 2.4--2.7~$\mu$m, while the broad water vapor band at 4.9 --7.7~$\mu$m overlaps with narrower signals of NO (5.0--5.7~$\mu$m), HNO$_3$ (4.7--6.0~$\mu$m), and NO$_2$ (6.0--6.4~$\mu$m).

However, due to the impact of the Carrington-like event, the reduction of H$_2$O reduces the atmospheric transit depth in all water bands, with the most significant reduction of up to 15$\%$ between 5.5 -- 7.0 $\mu$m (see the fourth panel of Fig.~\ref{fig:13}). In particular, the CO$_2$-poor runs [1,3,5] show a larger difference in the spectral features of H$_2$O than discussed in \cite{Wunderlich-etal-2020}. While \cite{Wunderlich-etal-2020} found a difference between scenarios [1] and [5] in the order of 12 ppm, this study shows a difference of 42 ppm, which directly can be attributed to the additional influence of ion chemistry in the upper atmosphere induced by the stellar event. Of further importance within the 5.0 -- 7.0 $\mu$m H$_2$O band are the two significant peaks around 5.2 -- 5.6~$\mu$m (NO) and 6.0 -- 6.5 $\mu$m (NO$_2$) showing deviations from the reference scenario of up to 40\%  (fourth panel of Fig.~\ref{fig:13}). While the increase in NO is mainly observed in the CO$_2$-poor wet-alive scenario [5], the increase in NO$_2$ is present in all scenarios. The latter is consistent with the formation of NO$_2$ up to 0.02~hPa (65~km) but is most vital in the two wet-alive scenarios [5,6]. These two scenarios further show an increase in the HNO$_3$ signal at 5.5 -- 6.0 $\mu$m. Here, the combined impact of the increase of the NO, HNO$_3$, and NO$_2$ signals masks the decrease of the water band signal in both scenarios. 

A decrease of up to 10$\%$ is shown in the methane band at 3.0 -- 3.7 $\mu$m, consistent with the decrease of methane due to the increased OH. However, loss of ozone could contribute to this as well, particularly the narrow features at $\sim$~3.3~$\mu$m and 3.6 -- 3.7~$\mu$m. A small positive feature in 3.3 -- 3.5 $\mu$m overlapping with the methane band can be attributed to another smaller NO$_2$ band, emphasizing the substantial increase of NO$_2$ particularly in the CO$_2$-rich dry-dead scenario [2], and in the two wet-alive scenarios [5, 6]. Conversely, the methane band at 7.0 -- 8.5~$\mu$m appears to increase. This is due to the increase of the HNO$_3$ band (at 7.3 -- 7.8~$\mu$m), which nearly perfectly overlaps with the maximum of the CH$_4$ band, shadowing the methane decrease in the spectrum. The increase of HNO$_3$ is also highlighted by the increase of the HNO$_3$ band at 11.0 -- 12.0 $\mu$m, which reaches up to 25$\%$ for the CO$_2$-poor wet-alive scenario [5]. The O$_3$ band at 9.0 -- 10.0 $\mu$m is reduced by 10 to 20$\%$ in all scenarios, most strongly for the wet-alive scenarios [5,6] with the highest oxygen amounts, and therefore O$_3$ values, at the beginning of the model period. This is counter-intuitive since we calculate strong increases in ozone at certain altitudes, and even in total ozone in some scenarios. This is because the ozone signals derive from the middle atmosphere above 20~km altitude as indicated by the effective height shown in Fig.~\ref{fig:franz}. In contrast, the total ozone column is more sensitive to the atmospheric parts below 100~hPa (20~km). 
\begin{figure}[!t]
    \centering
        \centering
        \includegraphics[width=0.7\columnwidth]{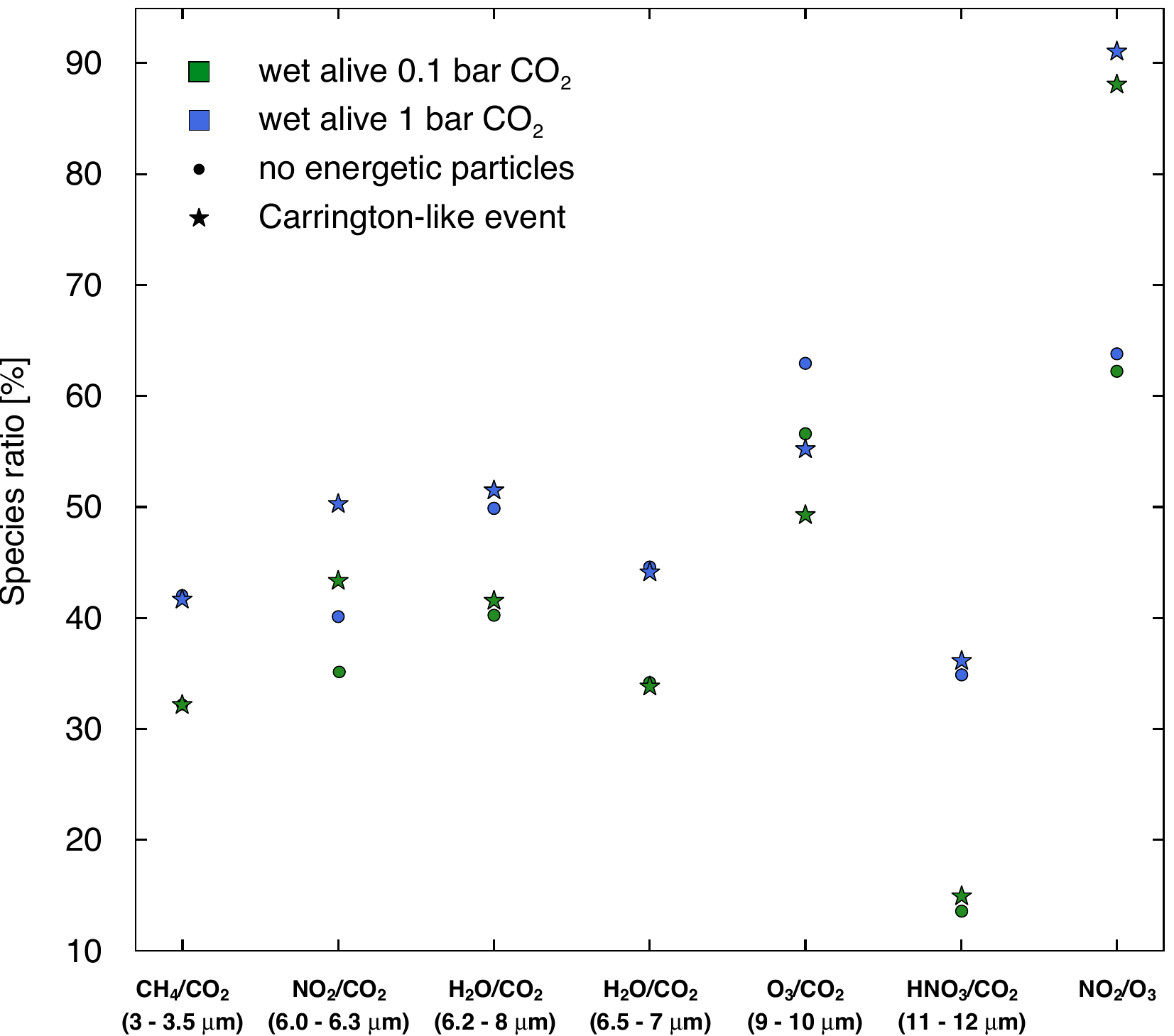}
    \caption{Ratio of individual spectral bands with respect to the CO$_2$ feature from 4.2 to 4.5 $\mu$m for the two Wet alive scenarios [5,6] with the greatest differences in potential biosignatures compared between Carrington-like event and no energetic particle event.
    }
    \label{fig:Ratios}
\end{figure}

For a clearer characterization of potentially measurable signatures Fig.~\ref{fig:Ratios} shows the ratios of individual spectral bands to the CO$_2$ band between 4.2 and 4.5 $\mu$m. CO$_2$ remains almost constant in all investigated scenarios and is therefore well suited as a reference band. When comparing with the CO$_2$ band, the most significant differences between model runs with and without particle event are shown for the NO$_2$ to CO$_2$ and O$_3$ to CO$_2$ ratios, while for CH$_4$ and HNO$_3$ there are no significant differences in this wavelength range. The selection of band boundaries also plays a role; an example is the water band, where for band boundaries from 6.2 -- 8.0 $\mu$m, including the whole band, the ratio increases for the model results with particle event, while for the more narrow selection of 6.5 -- 7.0 $\mu$m, the ratio decreases, which is caused by a contamination with NO$_2$ at 6.2 -- 6.3 $\mu$m in the first case. In Earth-like atmospheres, with a high proportion of N$_2$ and O$_2$, the ratio of NO$_2$/O$_3$ shows a large difference between cases with and without particle events and could, therefore, be used to characterize observations in the vicinity of a flaring star. Detecting such small signals, however, would be very challenging for the JWST.
\begin{figure*}[!t]
    \centering
    \includegraphics[width=\textwidth]{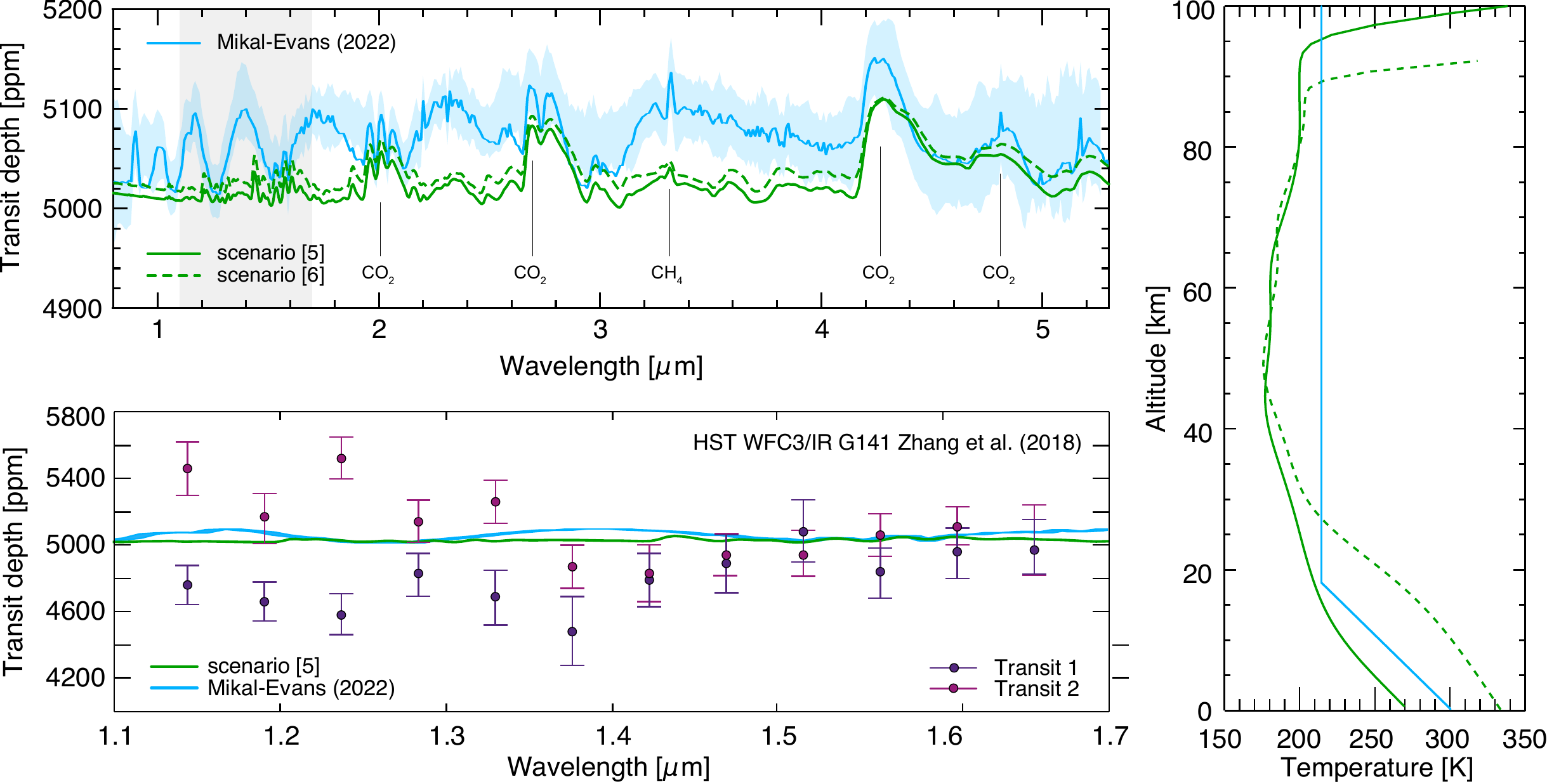}
    \caption{Upper left panel: Comparison of our work to the model results of \citet{mikal-evans_2022} who assumed a cloud-less Archean-Earth. Grey shading indicated the wavelength range of the figure below. Lower left panel: The transmission spectra of TRAPPIST-1e observed with HST/WFC3 during two different transits \citep[see][]{Zhang-etal-2018}. The green lines show our wet alive model results during the Carrington-like event (solid lines - scenario [5], dashed lines - scenario [6]), while the blue lines correspond to the Archean-Earth scenario of \citet{mikal-evans_2022}. Right panel: Comparison of the corresponding assumed atmospheric temperature profiles.}
   \label{fig:10}
\end{figure*}
\subsection{Comparison to Previous Model Efforts}
A recent study by \citet{mikal-evans_2022} discussed the possibility of observing the potential biosignature redox pair CH$_4$-CO$_2$ in the atmosphere of TRAPPIST-1e with the help of JWST. Utilizing the publicly available petitRADTRANS code \citep{Molliere-etal-2019}, transmission spectra for an Archean-like TRAPPIST-1e were derived. \citet{mikal-evans_2022} concluded that under the premise of reliable instrument noise and disregarding the impact of stellar variability, robust (5$\sigma$) detections of both CH$_4$ and CO$_2$ for 5 to 10 transit observations are feasible. The upper left panel of Fig.~\ref{fig:10} compares the cloud-free model results by \citet{mikal-evans_2022} (in blue) and our wet and alive scenarios, neglecting the impact of energetic particles. In contrast, the right panel compares the scenario-dependent temperature profiles. As expected, the results vary due to the different assumptions (i.e., Archean Earth versus modern Earth scenarios). However, the modeled CO$_2$ features agree and align well with the derived JWST/NIRSpec transmission spectra simulations (blue shaded area), assuming ten transits.

\subsection{Comparison to HST Observations}
In a study by \citet{Zhang-etal-2018}, the near-infrared transmission spectra (1.1 and 1.7 $\mu$m) of TRAPPIST-1e observed with the Hubble Space Telescope (HST) Wide Field Camera 3 (WFC3) during two transits - separated by twelve days - were discussed. As shown in the lower left panel of Fig.~\ref{fig:10}, above 1.4 $\mu$m, no spectral differences between the two transits can be seen. In the wavelength range $<$ 1.35 $\mu$m, however, differences of up to 650 ppm are present. One possible explanation was a "temporal variability of stellar contamination" \citep{Zhang-etal-2018}. Assuming the studied atmospheres to be valid representations, we find a good agreement between the observations and our model results - shown here as an example is the transition spectrum of the wet-alive 0.1 bar CO$_2$ scenario (solid green line) - at wavelengths above 1.35 $\mu$m. The same applies to the Archean-Earth scenario discussed in \citet{mikal-evans_2022} (blue line). However, none of our scenarios show a cosmic-ray-induced variation below 1.35 $\mu$m and thus rule out contamination of the transmission spectrum due to a Carrington-like event, as discussed in this study. However, more advanced studies, e.g., including variable exoplanetary magnetic fields and their response to potential coronal mass ejections passing the planet, are mandatory in the future.

\section{Summary and Conclusions}
With the help of our model suite INCREASE \citep[][]{Herbst-etal-2019c, Herbst-etal-2021}, we investigated the impact of cosmic rays (i.e., GCRs and StEPS) on atmospheric ionization, radiation exposure, ion chemistry, and - with that - the spectral transmission features of TRAPPIST-1e. Since its atmospheric composition is yet to be confirmed, we used six scenarios varying between dry-dead, wet-dead, and wet-alive conditions, assuming either CO$_2$-poor (0.1 bar) or CO$_2$-rich (1 bar) scenarios \citep[see][]{Wunderlich-etal-2020}.

Due to the lack of in situ particle measurements, we, for the first time, utilize a 1D SDE code to derive GCR proton and helium fluxes at the location of TRAPPIST-1e, showing the particle spectra to be comparable to spectra measured at Earth during very low solar activity phases. Therefore, GCRs form a non-neglectable production background concerning atmospheric ionization and radiation exposure, and with that, cannot be neglected in the context of (ion) chemistry, climate, and habitability. Furthermore, to study the impact of stellar energetic particles, the famous Carrington event has been scaled to the orbit of TRAPPIST-1e \citep[see discussion in ][]{Vida-etal-2017} and extended to energies up to 40 GeV. Because of the resulting high particle fluxes, we showed that the event-induced scenario-dependent atmospheric ionization and radiation exposure can be up to seven orders of magnitude higher than the GCR-induced values at the ionization maximum (see Figs.~\ref{fig:6} and \ref{fig:7}). 

As shown in Fig.~\ref{fig:14}, an enhancement of the ionization rates drastically impacts atmospheric ion chemistry. We found that for all six scenarios, the NOx production is comparable to terrestrial values induced by strong particle events. We further found CRs to strongly impact ozone in all investigated scenarios. For instance, for the CO$_2$-rich wet-dead scenario [4], ozone loss can be observed from 200~hPa (10 km) onward, while in the CO$_2$-poor wet-alive scenario [5] ozone production is observed at altitudes below 10~hPa (40 km). We found that the CO$_2$-poor dry-dead scenario [1] and the CO$_2$-poor wet-alive scenario [5], in particular at the beginning of the event, differ from the other scenarios by showing an increase in the total ozone column density (i.e., around 100 -- 10~hPa (15 - 40 km)). Further, we found that - in contrast to the other scenarios - the total ozone column density of scenarios [1] and [6] does not recover to background values within a few days after a Carrington-like event, which could strongly impact the transmission spectra of Earth-like exoplanets around active M-dwarfs with frequent flaring \citep[see also][]{Herbst-etal-2019c, Scheucher-etal-2020a}. We also showed that a Carrington-like event would lead to the destruction of water vapor and the production of HOx throughout the entire atmosphere and that hydroxyl acts as a methane sink in the wet-dead scenarios [3,4] (most prominent features are found within 90 -- 10~hPa (20 -- 50 km)) and the wet-alive scenarios [5,6] (most prominently around 10 -- 0.05~hPa (50 -- 70 km)). 

This work further illustrates the challenges of forthcoming transmission spectral analysis and the caution with which these spectra must be interpreted, especially in the case of Earth-like exoplanets orbiting active stars. In the scenario-dependent GARLIC-based synthetic planetary transmission spectra (Figs.~\ref{fig:TDcomp} and \ref{fig:13}) we found the following key features:
\begin{itemize}
\item[(1)] All scenarios show large event-induced amounts of NO$_2$ at 6.2 $\mu$m, most visible in scenarios [5,6] with an increase of 40\% compared to the reference case neglecting the impact of energetic particles.
\item[(2)] All scenarios show an event-induced reduction of the atmospheric transit depth in all water bands, most prominently seen for the H$_2$O band between 5.5 -- 7.0 $\mu$m with a decrease of up to 15\% in all six scenarios.
\item[(3)] Because of the simulated Carrington-like event, a decrease of up to 10\% in the methane band at 3.0 - 3.5 $\mu$m is present in all scenarios. In the case of scenarios [2,5,6], however, this feature is masked by the increase of NO$_2$ at 3.3 -- 3.5 $\mu$m. At first sight, the CH$_4$ band at 7.0 $\mu$m appears to increase - this increase, however, is actually caused by a strong HNO$_3$ band induced by the stellar energetic particles, which nearly perfectly overlaps with the spectral range of the CH$_4$ band, shadowing the methane decrease in the spectrum.
\item[(4)] Ozone features are rather weak for scenarios around mid-to-late M-dwarfs \citep[e.g., ][]{Scheucher-etal-2020a}. This is mainly due to changes in the incoming stellar radiation leading to a reduction in ozone abundance and spectral absorption \citep[e.g., ][]{Segura-etal-2005, Grenfell-etal-2012}. All scenarios show a depletion of ozone at 9.0 -- 10.0 $\mu$m in the order of 10 to 20$\%$. The amount of ozone loss is due to a sensitive balance between the formation of HOx, NOx, and atomic oxygen during the event, and therefore depends critically on the availability of atmospheric water vapor, $N_2$, and oxygen.
\item[(5)] The build-up of the HNO$_3$ spectral feature at 11.0 -- 12.0 $\mu$m, which previously has been reported to be a potential proxy of stellar particle contamination in all N$_2$-O$_2$-dominated atmospheres \citep[e.g., ][]{Tabataba-Vakili-2016, Scheucher-etal-2018, Herbst-etal-2019c, Scheucher-etal-2020a} is absent in the wet-dead scenarios [3,4].
\item[(6)] In the case of a CO$_2$-poor wet-alive Earth-like atmosphere, a strong event-induced NO feature at 5.5 $\mu$m is present. A similar feature can be found in the study by \citet{Scheucher-etal-2020a} investigating the impact of CRs on the Earth-like atmosphere of Proxima Centauri b.
\end{itemize}
\begin{figure}[!t]
    \centering
    \includegraphics[width=0.7\columnwidth]{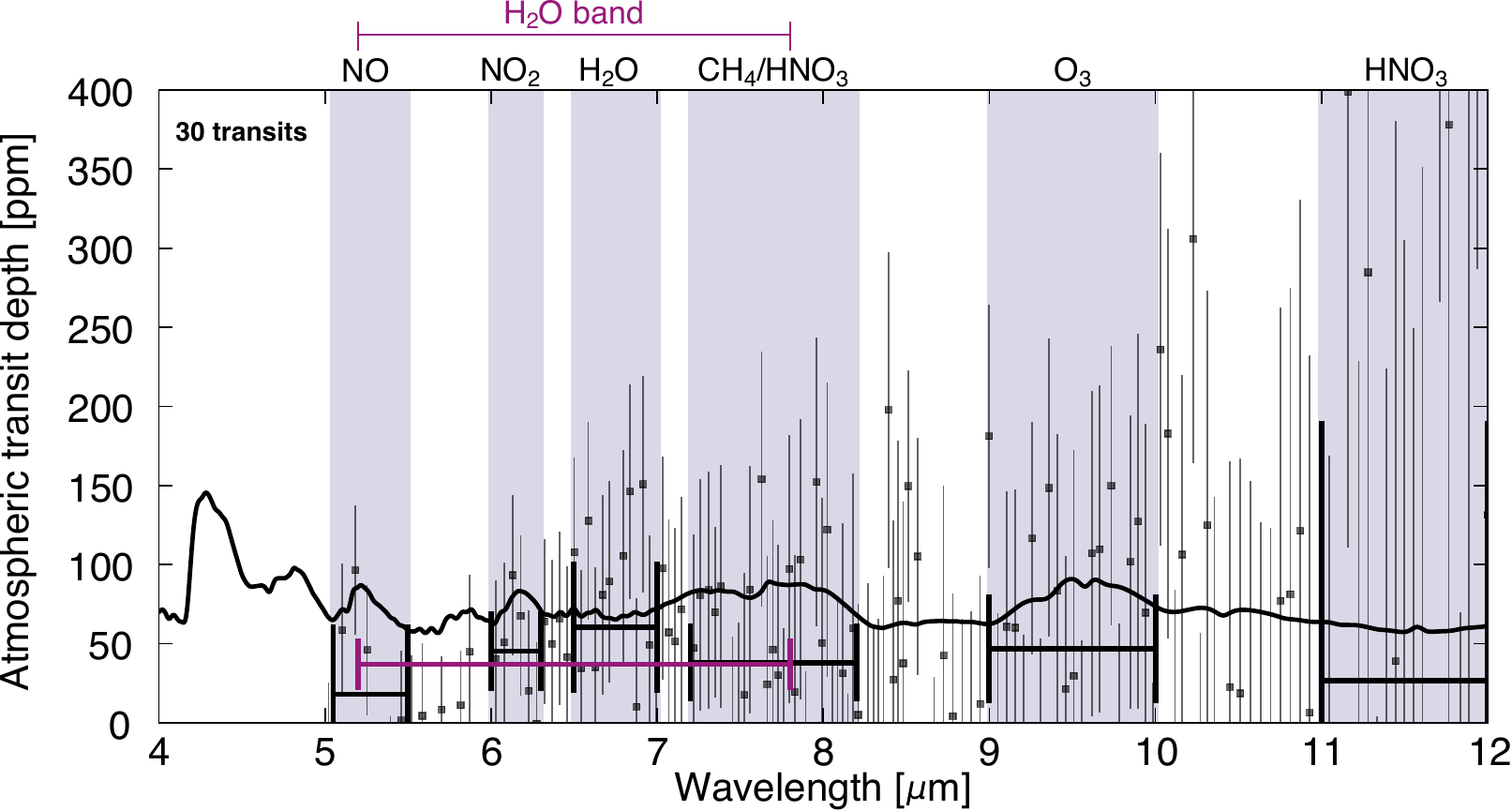}
    \caption{Modeled JWST  Mid-Infrared Instrument (MIRI) Low Resolution Spectroscopy (LRS) 30-transit transmission spectrum (dots) based on the CO$_2$-poor wet and alive scenario (solid black line).}
    \label{fig:MiriComparison}
\end{figure}
As discussed in \citet{Wunderlich-etal-2020}, the described scenarios may be differentiated under cloud-free circumstances by combining 30 transit observations with JWST NIRSpec. Using PandExo \citep{BatalhaEA2017}, we simulated JWST MIRI LRS observations of TRAPPIST-1e based on our wet and alive 0.1 bar CO$_2$ scenario during 1 to 100 transits (see Fig.~\ref{fig:MiriComparison} showing the results of the 30-transits run). In the case of the CO$_2$-poor wet and alive scenario, we find that to detect the O$_3$, NO$_2$, and H$_2$O features, 30 or more transits are necessary.

Unfortunately, even with 100 transits, the HNO$_3$ feature at 11 - 12 $\mu$m, which is a direct measure of the impact of cosmic rays on Earth-like atmospheres in our dry dead and wet alive scenarios [1,2,5,6], is not detectable with JWST MIRI due to the large signal-to-noise ratio in thermal IR. 
However, detecting such a feature is mandatory to distinguish whether the observed biosignature signals have been contaminated by the impact of stellar energetic particles. Therefore, future missions like the Origins Space Telescope (OST, see \href{https://origins.ipac.caltech.edu/download/MediaFile/171/original}{Mission Study Concept Report}), ELT \citep{NeichelEA2018}, and LIFE \citep{QuanzEA2022} are essential.

\section*{acknowledgements}
We acknowledge the support of the DFG priority program SPP 1992 “Exploring the Diversity of Extrasolar Planets (HE 8392/1-1, GR 2004/4-1, SI 1088/9-1, SM 486/2-1)”. KH is further grateful for the support of DFG grant 508335258 (HE 8392/2-1), acknowledges ISSI and the supported International Team 464 (ETERNAL), and would like to thank Prof. Dr. Bernd Heber and Dr. Sa\v{s}a Banjac (Christian-Albrechts-Universit\" at zu Kiel) for maintaining AtRIS. This work is based on the research supported partly by the National Research Foundation of South Africa (NRF grant number 137793).

\bibliography{references}{}
\bibliographystyle{aasjournal}

\end{document}